\documentclass[journal]{IEEEtran}
\usepackage{cite}
\usepackage{color,soul}
\usepackage{gensymb}
\usepackage{dsfont}
\ifCLASSINFOpdf
  \usepackage[pdftex]{graphicx}
  \graphicspath{{../pdf/}{../jpeg/}}
  \DeclareGraphicsExtensions{.pdf,.jpeg,.png}
\else
  \usepackage[dvips]{graphicx}
  \graphicspath{{../eps/}}
  \DeclareGraphicsExtensions{.eps}
\fi
\usepackage[cmex10]{amsmath}

\usepackage{amssymb}
\DeclareMathOperator*{\argmaxA}{arg\,max}{}
\DeclareMathOperator*{\argminA}{arg\,min}
\usepackage{euscript}

\usepackage{multirow}
\usepackage{algorithm}
\usepackage{algpseudocode}%[noend]
\usepackage{diagbox}
\usepackage{physics}
\usepackage{tikz}
\usepackage{comment}

\ifCLASSOPTIONcompsoc
  \usepackage[caption=false,font=normalsize,labelfont=sf,textfont=sf]{subfig}
\else
  \usepackage[caption=false,font=footnotesize]{subfig}
\fi
\usepackage{url}

\begin{document}
\title{Multi-Agent Safe Policy Learning for Power Management of Networked Microgrids}

\author{
         Qianzhi Zhang,~\IEEEmembership{Student Member,~IEEE,}
         Kaveh Dehghanpour,~\IEEEmembership{Member,~IEEE,}
         Zhaoyu Wang,~\IEEEmembership{Senior Member,~IEEE,} 
         Feng Qiu,~\IEEEmembership{Senior Member,~IEEE,}
         and Dongbo Zhao,~\IEEEmembership{Senior Member,~IEEE}
        % <-this % stops a space
\thanks{
This work was supported in part by the U.S. Department of Energy Wind Energy Technologies Office under Grant DE-EE0008956, and in part by the National Science Foundation under ECCS 1929975.

Q. Zhang, K. Dehghanpour, and Z. Wang are with the Department of
Electrical and Computer Engineering, Iowa State University, Ames,
IA 50011 USA (e-mail: wzy@iastate.edu).

F. Qiu and D. Zhao are with Energy Systems Division, Argonne National Laboratory,
Lemont, IL 60439 USA (e-mail: fqiu@anl.gov, dongbo.zhao@anl.gov).
}
}

\maketitle

\begin{abstract}
This paper presents a supervised multi-agent safe policy learning (SMAS-PL) method for optimal power management of networked microgrids (MGs) in distribution systems. While unconstrained reinforcement learning (RL) algorithms are \textit{black-box} decision models that could fail to satisfy grid operational constraints, our proposed method considers AC power flow equations and other operational limits. Accordingly, the training process employs the gradient information of operational constraints to ensure that the optimal control policy functions generate safe and feasible decisions. Furthermore, we have developed a distributed consensus-based optimization approach to train the agents' policy functions while maintaining MGs' privacy and data ownership boundaries. After training, the learned optimal policy functions can be safely used by the MGs to dispatch their local resources, without the need to solve a complex optimization problem from scratch. Numerical experiments have been devised to verify the performance of the proposed method. 
\end{abstract}


% \fi
\IEEEpeerreviewmaketitle
\section*{Nomenclature}
\addcontentsline{toc}{section}{Nomenclature}
\begin{IEEEdescription}[\IEEEusemathlabelsep\IEEEsetlabelwidth{$V_1,V_2,V_3$}]
\item[\textbf{Indices}]
\item[$i,j$] Indices of buses, $\forall i,j\in \Omega_I$.
\item[$ij$] Index of branch between bus $i$ and bus $j$, $\forall ij\in \Omega_{Br}$.
\item[$k$] Iteration index in distributed optimization, $k\in\{1,...,k^{max}\}$.
\item[$m$] Constraint index, $m\in\{1,...,M_c\}$.
\item[$n$] Agent index, $n\in\{1,...,N\}$.
\item[$t'$] Episode index in training process, $t' \in [t,t+T]$.

\item[\textbf{Parameters}]
\item[$a^f_n,b^f_n,c^f_n$] Coefficients of the DG quadratic cost function for agent $n$.
\item[$\pmb{b_{m,\mu_n}}$] Gradient vector of the constraint return function $m$ w.r.t. the parameters $\mu_n$.
\item[$\pmb{b_{m,\Sigma_n}}$] Gradient vectors of the constraint return function $m$ w.r.t. the parameters $\Sigma_n$.
\item[$D_n$] Dimension of multivariate Gaussian distribution function for agent $n$.
\item[$d_m$] Upper limit for constraint $m$.
\item[$E^{Cap}$] Max. capacity of ESS unit.
\item[$H_n$] Fisher information matrix of agent $n$.
\item[$\pmb{g_{\mu_n}}$] Gradient vector of the reward functions w.r.t. the parameters $\mu_n$.
\item[$\pmb{g_{\Sigma_n}}$] Gradient vector of the reward functions w.r.t. the parameters $\Sigma_n$.
\item[$I_{ij}^{M}$] Max. current limit on branch $ij$.
\item[$M_c$] Number of constraints.
\item[$M_c^G$] Number of global constraints.
\item[$M_c^L$] Number of local constraints.
\item[$N$] Number of MGs.
\item[$N_n$] Number of neighboring MGs for agent $n$.
\item[$P^{Ch,M}$] Max. ESS charging limits.
\item[$P^{Dis,M}$] Max. ESS discharging limits.
\item[$P^{D},Q^{D}$] Active and reactive load power.
\item[$P^{DG,M}$] Max. DG active power capacity.
\item[$Q^{DG,M}$] Max. DG reactive power capacity.
\item[$P^{DG,R}$] Max. DG ramp limit.
\item[$P^{PV}$] PV active power output.
\item[$P^{PCC,M}$] Max. active power flow at the PCCs.
\item[$Q^{PCC,M}$] Max. reactive power flow at the PCCs.
\item[$Q^{PV,M}$] Max. PV reactive power output limit.
\item[$SOC^{M}$] Max. SOC limits.
\item[$SOC^{m}$] Min. SOC limits.
\item[$T$] Length of the moving decision window.
\item[$V_{i}^{M},V_{i}^{m}$] Max. and min. voltage limit on bus $i$.
\item[$w_n(n')$] Weight parameters assigned of agent $n$ to neighboring agent $n'$.
\item[$Y^{Re},Y^{Im}$] Real and imaginary parts of the nodal admittance matrix $Y$.
\item[$\eta_{Ch},\eta_{Dis}$] Charging and discharging efficiency of ESS.
\item[$\lambda^{F}$] Diesel generator fuel price.
\item[$\lambda^R$] Retail price signals at the PCCs.
\item[$\pmb{\theta_{\mu_n}},\pmb{\theta_{\Sigma_n}}$] Vector of DNN weights and bias of agent $n$.
\item[$\pmb{\mu_n},\Sigma_n$] Mean vector and covariance matrices for control action of agent $n$.
\item[$\delta,\rho_1$] Step sizes for updating $\theta$ and $\lambda$.
\item[$\rho_2$] Penalty factor for constraints violation.
\item[$\Delta t$] Time step.
\item[$\Delta\theta_n$] Threshold for parameter updating.
\item[$\gamma$] Discount factor.
\item[$\tau$] Tightening multiplier.

\item[\textbf{Variables}]
\item[$\pmb{a_n}$] Vector of control actions of agent $n$.
\item[$C_m(\pi)$] Return value of constraint $m$ based on the control policy $\pi$.
\item[$F_{i,n}$] Fuel consumption of DG at bus $i$ of agent $n$.
\item[$I^{Re}_{i},I^{Im}_{i}$] Real and imaginary parts of the injected current at bus $i$.
\item[$I^{Re}_{ij},I^{Im}_{ij}$] Real and imaginary parts of the branch current at branch $ij$.
\item[$\pmb{O_t}$] Vectors of observation variable.
\item[$P^{Ch},P^{Dis}$] Charging and discharging power of ESS unit.
\item[$P^{DG},Q^{DG}$] DG active and reactive power outputs
\item[$P^{PCC}$] Active power flow at the PCC.
\item[$Q^{PCC}$] Reactive power flow at the PCC.
\item[$Q^{ESS}$] Reactive power outputs of ESS unit.
\item[$Q^{PV}$] PV inverter reactive power output.
\item[$SOC$] SOC of the battery system.
\item[$\pmb{S_n}$] Vectors of system state of agent $n$.
\item[$V^{Re}_{i},V^{Im}_{i}$] Real and imaginary parts of the bus voltage magnitude at bus $i$.
\item[$\pmb{\lambda_n}$] Vector of Lagrangian multipliers.

\item[\textbf{Functions}]
\item[$J_{R_{n}}$] Expected reward function of agent $n$.
\item[$J_{C_{m}}$] Expected return function of constraint $m$.
\item[$\pi_n$] Multivariate distribution function over control actions of agent $n$.
\item[$\Delta$] Kullback Leibler (KL)-divergence function.
\end{IEEEdescription}

\section{Introduction}
\IEEEPARstart{M}ICROGRIDS (MGs) are active clusters of distributed energy resources (DERs), loads, energy storage system (ESS),
and other onsite electric components. A smart distribution system may consist of multiple MGs and the coordinated control of the networked MGs can offer various benefits, including higher perpetration of local DERs, improved controllability, and enhancement of power system resilience and reliability \cite{Net_MGs,NM2018}. Solving the power management problem of networked MGs is a complex task. While previous works in this area have provided valuable insight, we have identified two shortcomings in the literature: 

(1) \textit{Limitations of model-based optimization methods:} In the existing literature, there are quite a few model-based methods for solving the optimal power management problem of networked MGs, such as centralized decision models \cite{Cen_MGs_2019,MMG_ZY,Liao_1} and distributed control frameworks \cite{Stochastic_2016_ZWang, Dis_dshi,Den_Bzhao}. However, with increasing number of MGs in distribution networks, these methods have to solve large-scale optimization problems with numerous nonlinear constraints that incur high computational costs and hinder real-time decision making. Furthermore, model-based methods are unable to adapt to the continuously evolving system conditions, as they need to re-solve the problem at each time step.

(2) \textit{Potential infeasibility of model-free machine learning methods:} To address the limitations of model-based methods, model-free reinforcement learning (RL) techniques have been used to solve the optimal power management problem through repeated interactions between a control agent and its environment. This approach eliminates the need to solve a large-scale optimization problem at each time point and enables the control agent to provide adaptive response to time-varying system states. Existing examples of RL application in power systems include economic dispatch and energy consumption scheduling of individual MGs \cite{RL_JDuan,DL_2018_PZeng,RL_2016_BKim} and multi-area smart control of generation in interconnected power grids \cite{RL_MG_TYU,Liao_2}. Further, in our previous paper \cite{QZ_RL}, we have proposed a bi-level power management method for networked MGs, where a centralized RL agent determines retail prices in a cooperative business model for each MG under the incomplete information of physical model. Current RL-based solutions employ control agents to train \textit{black-box} functions to approximate the optimal actions through trial and error. However, the trained black-box functions can fail to satisfy critical operational constraints, such as network nodal voltage and capacity limits, since these constraints have not been encoded in the training process. This can lead to unsafe operational states and control action infeasibility.

However, incorporating constraints into the training process of conventional black-box methods is challenging since these methods have generally relied on adding penalty terms to training objective functions for enforcing constraints, which cannot guarantee the safety of control policies as the number of constraints grows. Inspired by recent advances in constrained policy learning (PL) \cite{ConPolOpt2017, NIPS_2018_WM,  NIPS_2018_YC} and to address the shortcomings in the existing literature, we have cast the power management of networked MGs as a \textit{supervised multi-agent safe PL problem} (SMAS-PL). The various resources inside each MG and the collaborative behavior of MGs are both controlled to optimize the total cost of operation, while satisfying all the local and global constraints. Moreover, we have proposed a multi-agent policy gradient solution strategy, which enables individual MGs learn control policy functions to maximize the social welfare and ensure safety in a distributed way. The proposed method introduces a trade-off between model-free and model-based methods and combines the benefits offered by both sides. The purpose is to leverage the advantages of both model-free and model-based methods, for scalable real-time decision making while also maintaining a user-defined level of safety by considering constraints in the training process. Hence, on one hand, MGs' power management policy functions are modeled using black-box deep neural networks (DNNs); while on the other hand, to ensure decision feasibility, a constrained gradient-based training method is proposed that exploits the derivatives of the constraints and objective functions of the power management problem w.r.t. control actions and learning parameters. The training process employs these gradient factors to provide a convex quadratically constrained linear program (QCLP) approximation to the power management problem at each episode. This enables the proposed method to be both adaptable to changes in the inputs of the black-box components, and feasible with respect to operational constraints, including AC power flow. Finally, a distributed consensus-based primal-dual optimization method \cite{DisOpt_TH} is adopted to decompose the training task among MG agents. In summary, compared to existing decision making solutions, the main advantages of this paper are as follows:  
\begin{itemize}
\item Compared to the black-box learning-based methods, the proposed SMAS-PL leverages the gradient information of all the operational constraints to devise a tractable QCLP-based training process to promote the safety and feasibility of control policies. A backtracking mechanism is added into the PL framework to perform a final verification of feasibility before issuing control commands to the assets. 
\item Compared to conventional centralized training methods, the distributed training process in the SMAS-PL offers two advantages: it preserves the privacy of MG agents, including their control policies parameters and structures, operation cost functions, and local asset constraints; it also enhances computational efficiency and maintains scalability as the number of learning parameters grows into a humongous size.
\item The proposed SMAS-PL method does not need to solve a complex optimization problem in real-time. The agents' policy functions, that are trained offline, can be leveraged online to select optimal control actions in response to latest system state data.
\end{itemize}

The reminder of the paper is organized as follows: Section \ref{sec:framework} presents the overall framework of the proposed solution. Section \ref{sec:MGs} introduces the SMAS-PL problem and integrates problem gradients into the solver. Section \ref{sec:DistOpt} describes the multi-agent consensus-based training algorithm for SMAS-PL. Simulation results and conclusions are given in Section \ref{sec:Results} and Section \ref{sec:Con}, respectively. 
	
\section{Overview of the Proposed Framework}\label{sec:framework}
\begin{figure}
	\vspace{-0pt} 
	\vspace{-0pt}
	\centering
	\includegraphics[width=1.0\linewidth]{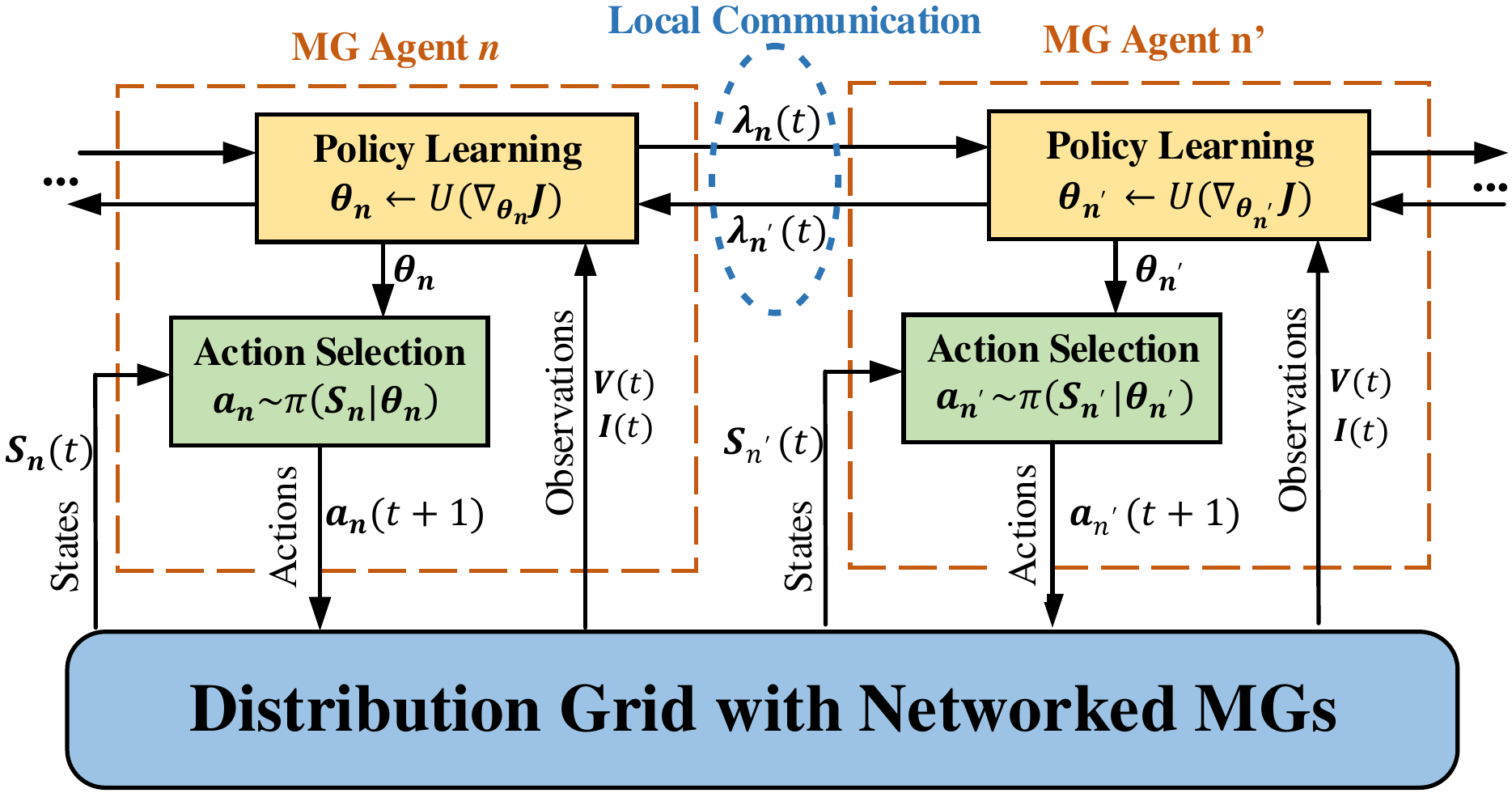}
	\vspace{-5pt} 
	\caption{Structure of the proposed SMAS-PL method for power management of networked MGs}
	\centering
	\label{framework}
    \vspace{-5pt} 
\end{figure}

The general framework of the proposed SMAS-PL method is shown in Fig. \ref{framework}. Note that vectors are denoted in bold letters throughout the paper. The micro-sources within each MG are controlled by an agent that adopts a private control policy. Here, the \textit{control policy} for the $n$'th agent, $\pi_n$, is a parametric probability distribution function, with parameters $\pmb{\theta_n}$, over the agent's control actions ($\pmb{a_{n,t}}$), including active/reactive power dispatching signals for local diesel generators (DGs), ESS and solar photo-voltaic (PV) panels. 
Note that the control policy $\pi_n$ is a function of the MG's state variables ($\pmb{S_{n,t}}$), defined by the aggregate MG load and solar irradiance. To ensure the safety of the control policies, MG agents receive the observed variables from the grid, including network nodal voltages $\pmb{V_t}$ and injection currents $\pmb{I_t}$, to determine gradient factors of the problem constraints and objectives w.r.t. to learning parameters, $\nabla_{\pmb{\theta}}\pmb{J}$. These gradient factors are then integrated into a multi-agent constrained training algorithm, which employs local inter-MG communication to satisfy all global and local operational constraints through exchanging and processing dual Lagrangian variables, $\lambda$(t). The Lagrangian multipliers embody the interactions among the MGs and capture the impacts of MGs' decisions on each other. Theoretical analysis and numerical simulations are conducted to show that the proposed SMAS-PL method can minimize the MG agents' operational cost and satisfy operational constraints. Note that the proposed SMAS-PL is not a purely model-free approach, since the AC power flow equations are used to calculate gradient factors and ensure the decision feasibility when training the DNNs.

In this paper, the MGs are chosen to be collaborative, because the satisfaction of the global constraints (i.e., limits on nodal voltages and line flows) for the whole network needs coordination among all MGs. Since global constraints are impacted by the response of all the MGs, we have devised a collaborative policy learning to ensure that grid-wide operation remains safe. Specifically, the consensus-based training method leverages the Lagrange multipliers of the global constraints to coordinate the policy optimization of the MGs. Thus, each Lagrange multiplier serves as a penalty factor or a shadow price, which enforces safety in the data-driven procedure.

\section{Safe Policy Learning for Power Management of Networked MGs}\label{sec:MGs}
To facilitate the discussion, Section \ref{subsec:Cen_MGs} introduces a general power management formulation that is commonly used in literature \cite{MMG_ZY, Stochastic_2016_ZWang, QZ_RL}. Sections \ref{subsec:setup} defines each component of the proposed SMAS-PL. In Sections \ref{subsec:SPL} and \ref{subsec:Gradient}, we propose a tractable SMAS-PL method, employing the gradient factors of reward function and constraint return functions w.r.t. actions and learning parameters, to solve the power management of networked MGs. 

\subsection{Power Management Problem Statement}\label{subsec:Cen_MGs}
Each MG is assumed to have local DGs, ESS, solar PV panels and a number of loads. This optimization problem is solved over a moving look-ahead decision window $t' \in [t,t+T]$, using the latest estimations of solar and load power at different instants. Here, $n$ is the MG index ($n\in\{1,...,N\}$), $i$ and $j$ define the node numbers ($\forall i,j\in \Omega_i$), $ij$ defines the branch numbers ($\forall ij\in \Omega_{Br}$).

\textit{1) \textbf{Problem objective}}: The objective function \eqref{eq3_1}, with control action vector $[P^{DG},P^{Ch},P^{Dis},Q^{DG},Q^{PV},Q^{ESS}]\in(\pmb{x_p},\pmb{x_q})$, minimizes MGs' total cost of operation, which is composed of the income/cost from power transfer with the grid and cost of running local DG. Here, $\lambda^{F}_{n}$ is the DG fuel price, $\lambda^{R}_{n}$ is the electricity price, and $P^{PCC}_{n,t'}$ is active power transfer between grid and the $n$'th MG at the point of common coupling (PCC). The fuel consumption of DG, $F_{i,n,t'}$, can be expressed as a quadratic polynomial function of its power, $P^{DG}_{i,n,t'}$, with parameters $a^{f}_{n}$, $b^{f}_{n}$ and $c^{f}_{n}$.
\begin{equation}
\min_{\pmb{x_p},\pmb{x_q}}\ \sum_{n=1}^{N}\sum_{t' = t}^{t+T}(-\lambda^{R}_{n}P^{PCC}_{n,t'}+\lambda^{F}_{i,n}F_{i,n,t'})\label{eq3_1}
\end{equation}
\begin{equation}
F_{i,n,t'}=a^{f}_n(P^{DG}_{i,n,t'})^2+b^{f}_nP^{DG}_{i,n,t'}+c^{f}_n\label{eq3_2}
\end{equation}

\textit{2) \textbf{Global constraints}}: These constraints are defined over variables that are impacted by control actions of all the MGs, including the voltage amplitude limits for the entire nodes, $[V_{i}^{m},V_{i}^{M}]$, and the maximum permissible branch current flow magnitudes $I_{ij}^{M}$ throughout the distribution grid and the MGs:
\begin{equation}
V_{i}^{m}\leq V_{i,t'}\leq V_{i}^{M}\label{eqPF_v}
\end{equation}
\begin{equation}
- I_{ij}^{M}\leq I_{ij,t'}\leq I_{ij}^{M}\label{eqPF_i}
\end{equation}

The global constraints \eqref{eqPF_v}-\eqref{eqPF_i} are implicitly determined by the AC power flow equations, which will be used to calculate the gradient factors of objective \eqref{eq3_1} and constraints \eqref{eqPF_v}-\eqref{eq3_essq} w.r.t. learning parameters as elaborated in Section \ref{subsec:Gradient}. Note that unlike previous centralized optimization solutions that are generally model-based, our strategy is a combination of both model-based and model-free approaches. Thus, while power flow equations appear explicitly in centralized optimization models, our solution only leverage power flow equations in an implicit way in the training process to ensure that the learning modules are generating feasible outcomes.

\textit{3) \textbf{Local constraints}}: These constraints are defined over the local control actions of each MG. Constraints \eqref{eq3_3}-\eqref{eq3_3q} ensure that the DG active/reactive power outputs, $P^{DG}_{i,n}/Q^{DG}_{i,n}$, are within the DG power capacity $P_{i,n}^{DG,M}/Q_{i,n}^{DG,M}$, and \eqref{eq3_4} enforces the maximum DG ramp limit, $P_{i,n}^{DG,R}$. PV reactive power output, $Q_{i,n}^{PV}$, is constrained by its maximum limit $Q_{i,n}^{PV,M}$ per \eqref{eq3_pv}. The active power transfer $P^{PCC}_{n,t'}$ and the reactive power transfer $Q^{PCC}_{n,t'}$ at the PCCs are bounded with the constraints \eqref{eqPCC} and \eqref{eqPCCQ}, respectively. 
\begin{equation}
0\leq P^{DG}_{i,n,t'}\leq P_{i,n}^{DG,M}\label{eq3_3}
\end{equation}
\begin{equation}
0\leq Q^{DG}_{i,n,t'}\leq Q_{i,n}^{DG,M}\label{eq3_3q}
\end{equation}
\begin{equation}
|P^{DG}_{i,n,t'}-P^{DG}_{i,n,t'-1}|\leq P_{i,n}^{DG,R}\label{eq3_4}
\end{equation}
\begin{equation}
|Q_{i,n,t'}^{PV}|\leq Q_{i,n}^{PV,M}\label{eq3_pv}
\end{equation}
\begin{equation}
|P^{PCC}_{n,t'}|\leq P_{n}^{PCC,M}\label{eqPCC}
\end{equation}
\begin{equation}
|Q^{PCC}_{n,t'}|\leq Q_{n}^{PCC,M}\label{eqPCCQ}
\end{equation}

The operational ESS constraints are described by \eqref{eq3_10}-\eqref{eq3_essq}, where \eqref{eq3_10} determines the state of charge (SOC) of ESSs, $SOC_{i,n}$. $E^{Cap}_{i,n}$ denotes the maximum capacity of ESSs. To ensure safe ESS operation, the SOC and charging/discharging power of ESS, $P^{Ch}_{i,n}$, $P^{Dis}_{i,n}$, are constrained as shown in \eqref{eq3_11}-\eqref{eq3_essq}. Here, $[SOC_{i,n}^{m},SOC_{i,n}^{M}]$, $P^{Ch,M}_{i,n}$ and $P^{Dis,M}_{i,n}$ define the permissible range of SOC, and maximum charging and discharging power, respectively. Constraint \eqref{eq3_14} indicates that ESSs cannot charge and discharge at the same time instant. And $\eta_{Ch}$/$\eta_{Dis}$ represents the charging/discharging efficiency. The reactive power of ESS, $Q_{i,n}^{ESS}$, is kept within maximum limit, $Q_{i,n}^{ESS,M}$, through constraint \eqref{eq3_essq}.
\begin{equation}
SOC_{i,n,t'} = SOC_{i,n,t'-1}+\Delta t \frac{(P^{Ch}_{i,n,t'}\eta_{Ch}- P^{Dis}_{i,n,t'}/\eta_{Dis})}{E^{Cap}_{i,n}}\label{eq3_10}
\end{equation}
\begin{equation}
SOC_{i,n}^{m}\leq SOC_{i,n,t'}\leq SOC_{i,n}^{M}\label{eq3_11}
\end{equation}
\begin{equation}
0\leq P^{Ch}_{i,n,t'}\leq P^{Ch,M}_{i,n}\label{eq3_12}
\end{equation}
\begin{equation}
0\leq P^{Dis}_{i,n,t'}\leq P^{Dis,M}_{i,n}\label{eq3_13}
\end{equation}
\begin{equation}
P^{Ch}_{i,n,t'}P^{Dis}_{i,n,t'} = 0\label{eq3_14}
\end{equation}
\begin{equation}
|Q_{i,n,t'}^{ESS}|\leq Q_{i,n}^{ESS,M}\label{eq3_essq}
\end{equation}

Note that the distribution system and networked MGs are operated in normal condition, which means the switch operation and the network topology are assumed to be unchanged during the operation period.

\subsection{Safe Policy Learning Setup}\label{subsec:setup}
In this section, the optimal power management of networked MGs is transformed into a SMAS-PL problem. The purpose of the SMAS-PL is to provide a framework for control agents to collaboratively find control policies to maximize their total accumulated reward while satisfying all problem constraints. To do this, we have provided formulations to ensure that the outcome of the SMAS-PL also corresponds to the solution of optimal power management of networked MGs \eqref{eq3_1}-\eqref{eq3_essq}. To show this, first we provide a description of the components of the SMAS-PL method:

\textit{1) \textbf{Control agents:}}
The problem consists of $N$ autonomous control agents, where each agent is in charge of dispatching the resources within an individual MG. The MGs are \textit{collaborative}, in the sense that they depend on local communication with each other to optimize their behaviors.

\textit{2) \textbf{State set:}}
The state vector for the $n$'th MG agent at time $t$ is defined as $\pmb{S_{n,t}}$ over the time window $[t,t+T]$, as $\pmb{S_{n,t}} = [\pmb{\hat{I}^{PV}_{n,t'}},\pmb{\hat{P}^{D}_{n,t'}}]^{t+T}_{t'=t}$, where $\pmb{\hat{I}^{PV}_{n,t'}}$ and $\pmb{\hat{P}^{D}_{n,t'}}$ are the vectors of predicted aggregate internal load power and solar irradiance of the $n$'th MG at time $t'$, respectively. The prediction errors follow random distributions with zero mean and the standard deviations selected from the beta and Gaussian distributions adopted from \cite{Solar_beta,Solar_beta2,Load_Wu}. Note that the parameters of forecasting error distributions are different for different MG agents.

\textit{3) \textbf{Action Set:}}
The control action vector for the $n$'th agent at time $t$ is denoted as $\pmb{a_{n,t}} \in \mathbb{R}^{D_n}$ and consists of the dispatching decision variables for the $n$'th MG over the time window $[t,t+T]$, as $\pmb{a_{n,t}} = [P^{DG}_{n,t'},P^{Ch}_{n,t'},P^{Dis}_{n,t'},Q^{DG}_{n,t'},Q^{PV}_{n,t'},Q^{ESS}_{n,t'}]^{t+T}_{t'=t}$.

\textit{4) \textbf{Observation Set:}}
The observation variable vector for the agents at time $t$ is denoted as $\pmb{O_t}$, and includes grid's nodal voltages $\pmb{V_t}$ and current injections $\pmb{I_t}$ at that time, $\pmb{O_t}=[\pmb{V_t},\pmb{I_t}]$. Note that the observations are implicitly determined by the agents' control actions, and thus, cannot be predicted independently of the agents' policies. However, unlike the observation variables, the state variables are independent of the agents' control actions and can be predicted for the whole decision window without the need to consider agents' policies. In the power management problem, nodal sensors or distribution grid's state estimation module will provide the latest values of observations.

\textit{5) \textbf{Control policy:}}
In this work, the control policies are modelled as multivariate Gaussian distributions due to several reasons: (i) Gaussian distributions allow for explicit learning of both expectations and uncertainties of control policies, which are directly represented by the parameters of the distribution. Most of other distributions are parameterized by unintuitive parameters that make the decision model harder to interpret and verify. (ii) The gradients of Gaussian policy functions with respect to actions and learning parameters are easy to compute (see Appendices A and B). (iii) Gaussian policy functions have been adopted and suggested by \cite{RL_2017_RSS} and \cite{Constrined_EV}. Thus, the control policy for the $n$'th agent, denoted as $\pi_n$, is defined as a $D_n$-dimensional multivariate Gaussian distribution over control actions $\pmb{a_n}$. The policy function determines the probability of the agent's optimal control action after training, as follows:
\begin{equation}
\pmb{a_n} \sim \pi_{n}(\pmb{a_n}|\pmb{\theta_n})=\frac{1}{\sqrt{|\Sigma_n|(2\pi)^{D_n}}}e^{-\frac{1}{2}(\pmb{a_n}-\pmb{\mu_n})^\top\Sigma_n^{-1}(\pmb{a_n}-\pmb{\mu_n})}\label{eq2_12_1}
\end{equation}
where $\pmb{\mu_n}\in \mathbb{R}^{D_n \times 1}$ is the mean vector and $\Sigma_n \in \mathbb{R}^{D_n \times D_n}$ is the covariance matrix of of multivariate Gaussian distribution for the $n$'th agent. The Gaussian policy function explicitly determines the expected value and uncertainties of optimal control actions for each agent. Each agent's learning parameter vector, $\pmb{\theta_{n}}$, consists of two parametric subsets $\pmb{\theta_{\mu_n}}$ and $\pmb{\theta_{\Sigma_n}}$, corresponding to the mean vector and the covariance matrix of the agent's policy function. To do this, two DNNs are used for each MG agent as parametric learning functions to represent control policy components. These DNNs receive the agent's states, $\pmb{S_{n}}$, as input to fully quantify the sufficient statistics of optimal control policies of MGs, i.e., the mean vector and the covariance matrix of the agent's actions, as follows:
\begin{equation}
\pmb{\mu}_n=DNN(\pmb{S_n}|\pmb{\theta_{\mu_n}})\label{DNN_1}
\end{equation}
\begin{equation}
\Sigma_n=DNN(\pmb{S_n}|\pmb{\theta_{\Sigma_n}})\label{DNN_2}
\end{equation}

The DNNs are maintained, continuously updated, and deployed in real-time by local control agents of each MG. Note that the proposed SMAS-PL method introduces a trade-off between model-free and model-based methods and combines the benefits offered by both sides. Thus, the reasons for the use of DNN-based distributions for modeling actions are as follows: (i) we have leveraged the model information to train safe policy functions that guarantee feasibility (i.e., the model-based aspect of the solution); (ii) the trained policy functions are deployed online for action selection, simply by inserting the latest data samples into the DNN-based policy functions (i.e., the model-free aspect of the solution).

\textit{6) \textbf{Reward function:}}
The reward function for the $n$'th MG is defined as the discounted negative accumulated operational cost of individual MG over the decision window $[t,t+T]$, $R_{n,t'}=-[\sum_{t'=t}^{t+T}(-\lambda^{R}_{n}P^{PCC}_{n,t'}+\lambda^{F}_{i,n}F_{i,n,t'})]$, obtained from the objective functions of the networked MGs power management problem, \eqref{eq3_1}, as follows:
\begin{equation}
J_{R_n}(\pi_n) = E_{\pi_n}[\sum_{t'=t}^{t+T}\gamma^{t'}R_{n,t'}], \forall n \in\{1,...,N\}\label{eq2_01}
\end{equation}
where, $\gamma\in[0,1)$ is a discount factor that determines each MG agent's bias towards rewards received at different time instances. An agent with $\gamma=0$ is a purely-myopic decision maker, which favors immediate reward at the expense of later expected reward values. On the other hand, $\gamma = 1$ represents an unbiased agent, which assigns equal weights to the reward received at all time instants. This parameter is user-defined and depends on each MG's economic priorities. The expectation operation $E_{\pi_n}\{\}$ is used to calculate reward with respect to the future expected action-states, which are in turn impacted by the uncertainties of states and observations.

\textit{7) \textbf{Constraint return:}}
The SMAS-PL consists of a total of $M$ constraints, including $M_c^L$ local and $M_c^G$ global constraints, defined by \eqref{eqPF_v}-\eqref{eqPF_i} and \eqref{eq3_3}-\eqref{eq3_essq}, respectively, and denoted as $C_m(\pi) \leq d_m, m \in\{1,...,M_c\}$, where $C_m(\pi)$ represents the return value of $m$'th constraint under the control policy $\pi$ and $d_m$ is the upper-boundary of the $m$'th constraint. Note that all constraints in the power management problem have been transformed into this format (equality constraint \eqref{eq3_14} can be transformed into two inequality constraints). Constraint satisfaction is encoded into the SMAS-PL using the discounted constraint return values of agents' policies $\pi$ as:
\begin{equation}
J_{C_m}(\pi) = E_{\pi}[\sum_{t'=t}^{t+T}\gamma^{t'}C_{m,t'}]\leq d_m, \forall m \in\{1,...,M_c\}\label{eq2_02}
\end{equation}
where, expectation operation has been leveraged in \eqref{eq2_02} to handle the state and observation uncertainties.

\subsection{Safe Policy Learning Formulation}\label{subsec:SPL}
Given the definitions of the components of the SMAS-PL (Section \ref{subsec:setup}), the power management problem of the networked MGs \eqref{eq3_1}-\eqref{eq3_essq} is transformed into an iterative SMAS-PL problem, where the control policies of the agents are updated at time $t$, around their latest values, by maximizing a reward function \eqref{eq2_1}, while satisfying constraint return criteria: 
\begin{equation}
\pmb{\pi}^{t+1} = \argmaxA_{ \pi_1,...,\pi_N}\sum_{n=1}^{N}J_{R_n}(\pi_n)\label{eq2_1}
\end{equation}
\begin{equation}
s.t.\ \ \pmb{a_{n}} \sim \pi_{n}(\pmb{S_n})\label{eq2_1_0}
\end{equation}
\begin{equation}
J_{C_{m}}(\pmb{\pi})\leq d_{m},\ \forall m\label{eq2_2}
\end{equation}
\begin{equation}
\Delta(\pi_n,\pi_n^{t}) \leq \delta,\ \forall n\label{eq2_3}
\end{equation}
where, $\pmb{\pi} = \{\pi_1,...,\pi_n\}$ denotes the set of control policies of all agents. In \eqref{eq2_1_0}, the agent's policy is a function of the state vector, $\pmb{S_n}$. In \eqref{eq2_2}, the expected constraint return value are used to ensure the satisfaction of $m$'th constraint based on control policies. In \eqref{eq2_3}, $\Delta(\cdot,\cdot)$ is the Kullback Leibler (KL)-divergence function \cite{ConPolOpt2017} that serves as a distance measure between the previous policy, $\pi_n^t$, and the updated policy, $\pi_n^{t+1}$, and is constrained by a step size, $\delta$. Note that \eqref{eq2_3} ensures that consecutive policies are within close distance from each other.  

The intractable non-convex PL formulation, \eqref{eq2_1}-\eqref{eq2_3}, can be solved in principle using a trust region policy optimization (TRPO) method \cite{ConPolOpt2017}; however, in this paper we apply a further approximation to TRPO to transform the problem into a tractable convex iterative QCLP, which enables learning the PL parameters, $\pmb{\theta}=\{\pmb{\theta_1},...,\pmb{\theta_N}\}$, in a more scalable and efficient manner. Our solution leverages the linear approximations of the objective and constraint returns around the latest parameter values $\pmb{\theta^{t}}$: 
\begin{equation}
\pmb{\theta^{t+1}} = \argmaxA_{\pmb{\theta_1},...,\pmb{\theta_N}}\sum_{n=1}^{N}\pmb{g_n}^{T}(\pmb{\theta_n}-\pmb{\theta_n^t})\label{eq2_7}
\end{equation}
\begin{equation}
s.t.\ \ J_{c_{m}}(\pmb{\theta^t})+\pmb{b_{m}}^{T}(\pmb{\theta}-\pmb{\theta^t})\leq d_{m},\ \forall m\label{eq2_8}
\end{equation}
\begin{equation}
\frac{1}{2}(\pmb{\theta_n}-\pmb{\theta_n^t})^{T}H_n(\pmb{\theta_n}-\pmb{\theta_n^t})\leq \delta,\ \forall n\label{eq2_9}
\end{equation}
where, $\pmb{g_n} = \nabla_{\theta} J_R$ and $\pmb{b_m} = \nabla_{\theta} J_{C_m}$ are the \textit{gradient factors} of the reward and constraint return functions w.r.t. the learning parameters. Constraint \eqref{eq2_3} is transformed into \eqref{eq2_9} using the Fisher information matrix (FIM) of the policy functions, $\pi_n$, denoted by $H_n$. The FIM is a positive semi-definite matrix, whose $(c,d)$'th entry for policy functions with a Gaussian structure is determined as follows \cite{FIM_2013}:
\begin{multline}\label{eq2_H}
H_n(c,d)=E[\frac{\partial \log \pi_n(\pmb{a_n}|\pmb{\theta_n})}{\partial \pmb{\theta_n}(c)}\frac{\partial \log \pi_n(\pmb{a_n}|\pmb{\theta_n})}{\partial \pmb{\theta_n}(d)}]\\
=2(\frac{\partial \mu^{H}_n}{\partial \pmb{\theta_n}(c)}\Sigma_n^{-1}\frac{\partial \pmb{\mu_n}}{\partial \pmb{\theta_n}(d)})+\Tr{\Sigma_n^{-1}\frac{\partial \Sigma_n}{\partial \pmb{\theta_n}(c)}\Sigma_n^{-1}\frac{\partial \Sigma_n}{\partial \pmb{\theta_n}(d)}}
\end{multline}

Note that \eqref{eq2_7}-\eqref{eq2_9} provides a convexified constrained gradient-based method for training the policy functions' parameters of the MG agents; using this QCLP-based strategy the agents do not need to learn an action-value function explicitly. Instead, the power-flow-based gradient factors, $\pmb{g_n}$ and $\pmb{b_m}$, have to be determined for the two sets of learning parameters, $[\pmb{\theta_{\mu_n}},\pmb{\theta_{\Sigma_n}}]$. This process is outlined in Section \ref{subsec:Gradient}. 

\subsection{Gradient Factor Determination}\label{subsec:Gradient}
To determine gradient factors, the following information are used: (i) the observation variables, $\pmb{O_t}$, including nodal voltage $\pmb{V}$ and current injections $\pmb{I}$; (ii) the latest system states $\pmb{S_{n,t}}$ for each MG agent; (iii) the latest control actions $\pmb{a_n}$ of each MG agent; (iv) the latest learning parameters $\pmb{\theta_n}=[\pmb{\theta_{\mu_n}},\pmb{\theta_{\Sigma_n}}]$; (v) network parameters, including the nodal admittance matrix, $Y$. Using information (i)-(v) and chain rule, $\pmb{g_n}=[\pmb{g_{\mu_n}},\pmb{g_{\Sigma_n}}]$ and $\pmb{b_{m}}=[\pmb{b_{m,\mu_n}},\pmb{b_{m,\Sigma_n}}]$ in \eqref{eq2_7} and \eqref{eq2_8} can be written as: 
\begin{subequations}\label{eq2_change_1}
\begin{equation}
\pmb{g_{\mu_n}}=\frac{\partial J_{R_n}}{\partial \pmb{a_n}}\frac{\partial \pmb{a_n}}{\partial \pi_n}\frac{\partial \pi_n}{\partial \pmb{\mu_n}}\frac{\partial \pmb{\mu_n}}{\partial \pmb{\theta_{\mu_n}}}
\end{equation}
\begin{equation}
\pmb{b_{m,\mu_n}}=\frac{\partial J_{C_m}}{\partial \pmb{a_n}}\frac{\partial \pmb{a_n}}{\partial \pi_n}\frac{\partial \pi_n}{\partial \pmb{\mu_n}}\frac{\partial \pmb{\mu_n}}{\partial \pmb{\theta_{\mu_n}}}
\end{equation}
\end{subequations}
\begin{subequations}\label{eq2_change_2}
\begin{equation}
\pmb{g_{\Sigma_n}}=\frac{\partial J_{R_n}}{\partial \pmb{a_n}}\frac{\partial \pmb{a_n}}{\partial \pi_n}\frac{\partial \pi_n}{\partial \Sigma_n}\frac{\partial \Sigma_n}{\partial \pmb{\theta_{\Sigma_n}}}
\end{equation}
\begin{equation}
\pmb{b_{m,\Sigma_n}}=\frac{\partial J_{C_m}}{\partial \pmb{a_n}}\frac{\partial \pmb{a_n}}{\partial \pi_n}\frac{\partial \pi_n}{\partial \Sigma_n}\frac{\partial \Sigma_n}{\partial \pmb{\theta_{\Sigma_n}}}
\end{equation}
\end{subequations}
where, each gradient factor, $\pmb{g_{\mu_n}}$, $\pmb{b_{m,\mu_n}}$, $\pmb{g_{\Sigma_n}}$, and $\pmb{b_{m,\Sigma_n}}$, consists of four elements. All the elements in \eqref{eq2_change_1} and \eqref{eq2_change_2} can be obtained as follows:

\textit{1) $\partial J_{R_n}/\partial \pmb{a_n}$ and $\partial J_{C_m}/\partial \pmb{a_n}$}: The gradients of the expected reward $J_{R_n}$ and the expected constraint return $J_{C_m}$ w.r.t. control actions $\pmb{a_n}$ can be obtained using a proposed four-step process, that leverages the current injection-based AC power flow equations. The details of this process are shown in Appendix \ref{app:gradient}. 

\textit{2) $\partial \pmb{a_n}/\partial \pi_n$}: Using the latest values for parameters $\pmb{\mu_n}$, $\Sigma_n$, and actions $\pmb{a_n}$, the gradient of control actions w.r.t. $\pi_n$ is obtained from \eqref{eq2_12_1}, as shown in \eqref{eq_step_2}:
\begin{equation}
\frac{\partial \pmb{a_n}}{\partial \pi_n}=-(\frac{\Sigma_n^{-1}(\pmb{a_n}-\pmb{\mu_n)}}{\sqrt{|\Sigma_n|(2\pi)^{D_n}}}e^{-\frac{1}{2}A})^{-1}\label{eq_step_2}
\end{equation}
where, $A=(\pmb{a_n}-\pmb{\mu_n})^\top\Sigma_n^{-1}(\pmb{a_n}-\pmb{\mu_n})$. The 
detailed derivation of \eqref{eq_step_2} can be found in Appendix \ref{app:proof}.

\textit{3) $\partial \pi_n/\partial \pmb{\mu_n}$ and $\partial \pi_n/\partial \Sigma_n$}: using the latest values for parameters $\pmb{\mu_n}$, $\Sigma_n$ and actions $\pmb{a_{n}}$, the gradients of control policies, w.r.t. $\pmb{\mu_n}$ and $\Sigma_n$ are determined using \eqref{eq2_12_1}, as shown in \eqref{eq_step_31} and \eqref{eq_step_32}:
\begin{equation}
\frac{\partial \pi_n}{\partial \pmb{\mu_n}}=\frac{\Sigma_n^{-1}(\pmb{a_n}-\pmb{\mu_n})}{\sqrt{|\Sigma_n|(2\pi)^{D_n}}}e^{-\frac{1}{2}A}\label{eq_step_31}
\end{equation}
\begin{equation}\label{eq_step_32}
\frac{\partial \pi_n}{\partial \Sigma_n}=-\frac{1}{2}\frac{(\Sigma_n^{-1}-\Sigma_n^{-1}(\pmb{a_n}-\pmb{\mu_n})(\pmb{a_n}-\pmb{\mu_n})^\top\Sigma_n^{-1})}{\sqrt{|\Sigma_n|(2\pi)^{D_n}}}e^{-\frac{1}{2}A}
\end{equation}
where, the detailed derivations of \eqref{eq_step_31} and \eqref{eq_step_32} are shown in Appendix \ref{app:proof}.

\textit{4) $\partial \pmb{\mu_n}/\partial \pmb{\theta_{\mu_n}}$ and $\partial \Sigma_n/\partial \pmb{\theta_{\Sigma_n}}$}: 
A \textit{back-propagation process} \cite{BP_2012_ANN} is performed on the two DNNs within each MG agent's control policy function, \eqref{DNN_1} and \eqref{DNN_2}, to determine the gradients of DNNs' outputs w.r.t. their parameters. In each iteration, the latest values of state variables are employed as inputs of the DNNs. The back-propagation process exploits chain rule for stage-by-stage spreading of gradient information through layers of the DNNs, starting from the output layer and moving towards the input \cite{BP_2012_ANN}. To enhance the stability of the back-propagation process, a sample batch approach is adopted, where the gradients obtained from several sampled actions are averaged to ensure robustness against outliers.    

\section{Multi-Agent Consensus-based Safe Policy Learning}\label{sec:DistOpt}
\subsection{Offline Policy Training}\label{subsec:offline}
Using the gradient factors \eqref{eq2_change_1} and \eqref{eq2_change_2}, the QCLP, \eqref{eq2_7}-\eqref{eq2_9}, is fully specified and can be solved at each policy update iteration for training the agents' PL frameworks. However, we have identified two challenges in this problem: (i) the size of the DNN parameters $\pmb{\theta}$ can be extremely large, which results in high computational costs during training; (ii) the control policy privacy of the MG agents needs to be preserved during training, which implies that the agents might not have access to each other's control policies, cost functions, and local constraints on assets. Centralized solvers can be both time-consuming and lack guarantees for maintaining data ownership boundaries. 

In order to address these two challenges, we have developed a \textit{multi-agent consensus-based constrained training algorithm} \cite{DisOpt_TH}. Due to its distributed nature this method is both scalable and does not require sharing control policy parameters among agents. Thus, the proposed algorithm is able to efficiently solve the QCLP \eqref{eq2_7}-\eqref{eq2_9}, while relying only on local inter-MG communication. The purpose of inter-MG interactions is to satisfy global constraints, \eqref{eqPF_v}-\eqref{eqPF_i}. To do this, the agents repeatedly estimate and communicate dual variable $\pmb{\lambda_n}$, corresponding to the Lagrangian multiplier of global constraints. Furthermore, a local primal-dual gradient step is included in the algorithm to move the primal and dual parameters towards their global optimum. The proposed distributed algorithm consists of four stages that are performed iteratively, as follows: 

\textbf{Stage I.} \textit{Initialize ($k\leftarrow 1$):} Gradient factors $\pmb{g_n}$ and $\pmb{b_m}$ are obtained from Section II-D. The previous values of learning parameters are input to the QCLP, $\pmb{\theta_n^t}(0)\leftarrow\pmb{\theta_n^{t-1}}$. Lagrangian multipliers are initialized as zero for each MG agent.

\textbf{Stage II.} \textit{Weighted averaging operation:} MG agent $n$ receives the Lagrangian multiplier $\pmb{\lambda_{n'}}$, for global constraints \eqref{eqPF_v}-\eqref{eqPF_i}, from its neighbouring MG agents $n'\in\{1,...,N_n\}$ and combines the received estimates using weighted averaging:
\begin{equation}
\pmb{\bar{\lambda}_n}(k)=\sum_{{n'}=1}^{N_n}w_{n}(n')\pmb{\lambda_{n'}}(k)\label{eq2_DOpt_1}
\end{equation}
where, $w_n(n')$ is the weight that MG agent $n$ assigns to the incoming message of the neighbouring MG agent $n'$. To guarantee convergence to consensus, the weight matrix, composed of the agents' weight parameters is selected as a doubly stochastic matrix \cite{DisOpt_TH}, i.e., $w_n(n') = \frac{1}{N_n}$. This weight selection strategy implies that the MG agents assign equal importance to the information received from their neighboring agents.

\textbf{Stage III.} \textit{Primal gradient update:} The $n$'th MG agent updates its primal parameters $\pmb{\theta_n^t}$ employing a gradient descent operation, using the gradients of the agent's reward and the global constraint returns, $m'\in M_c^G$, and step size $\rho_1$:
\begin{equation}\label{eq2_DOpt_2}
\pmb{\bar{\theta}_n}(k)= \pmb{\theta_n^t}(k)\\
-\rho_1(\pmb{g_n}(\pmb{\theta^t_n}(k))+\pmb{b_{m'}}(\pmb{\theta^t_n}(k))\pmb{\bar{\lambda}_n}(k))
\end{equation}

\textbf{Stage IV.} \textit{Projection on local constraints:} The agent projects the local learning parameters to the feasible region defined by the gradients of the local constraints \eqref{eq3_3}-\eqref{eq3_essq}:
\begin{equation}
\pmb{\theta_n^t}({k+1}) = \argminA_{\pmb{\theta}}||\pmb{\bar{\theta}_n}(k)-\pmb{\theta}||\label{eq2_DOpt_3}
\end{equation}
\begin{equation}
s.t.\ \ J_{c_{m}}(\pmb{\theta_n^t}(0))+\pmb{b_{m}}^{T}(\pmb{\theta^t_n}(0)-\pmb{\theta})\leq d_{m},\ \forall m \in M_c^L\label{eq2_DOpt_4}
\end{equation}
\begin{equation}
\frac{1}{2}(\pmb{\theta^t_n}(0)-\pmb{\theta})^{T}H_n(\pmb{\theta^t_n}(0)-\pmb{\theta})\leq \delta,\ \forall n\label{eq2_DOpt_5}
\end{equation}
 
\textbf{Stage V.} \textit{Dual gradient update:} Each agent's estimations of dual variables $\pmb{\lambda_n}$ for the global constraints, \eqref{eqPF_v} and \eqref{eqPF_i}, will be updated using a gradient ascent process over $\pmb{\bar{\lambda}_n}$:
\begin{equation}
\pmb{\lambda_n}(k+1) = [(\pmb{\bar{\lambda}_n}(k)+\rho_2(\pmb{b_{m'}}\pmb{\theta^t_n}(k+1)-d_{m'})]^+, \forall m' \in M_c^G\label{eq2_DOpt_3_1}
\end{equation}
where, $\rho_2$ is a penalty factor for global constraints violation, and the operator $[\cdot]^+$ returns the non-negative part of its input. 

\textbf{Stage VI.} \textit{Stopping criteria:} Check algorithm convergence using the changes of $\pmb{\theta^t_n}(k)$; stop when the changes in parameters falls below the threshold value $\Delta\theta_{n}$; otherwise, go back to Stage II. 

The overall flowchart of the SMAS-PL training process using the proposed distributed training technique is shown in Algorithm \ref{alg:DCPO}. The calculations of Steps 8 and 9 can be found in Appendix \ref{app:gradient}. 
\begin{algorithm}
\caption{SMAS-PL Training}\label{alg:DCPO}
\begin{algorithmic}[1]
\State {Select $t^{max}, T, \delta, k^{max}, w_{n}(n'), \rho_1, \rho_2, \Delta\theta_n$}
\State {Initialize $\pmb{\theta^{t_0}_n}$}
\For{$t \gets 1$ to $t^{max}$} 
        \State {$\pmb{S_n} \leftarrow [\pmb{S_n}(t), ..., \pmb{S_n}(t+T)]$}
        \State {$\pmb{\mu_n} \leftarrow$ \eqref{DNN_1} [Parameter insertion]}
        \State {$\pmb{\Sigma_n} \leftarrow$ \eqref{DNN_2} [Parameter insertion]}   
        \State {$\pmb{a_n} \sim \pi_{n}(\pmb{S_n}|\pmb{\theta_n}) \leftarrow \eqref{eq2_12_1}$ [Action selection]}
        \State{$\partial J_{R_n}/\partial \pmb{a_n} \leftarrow$ \eqref{eq_Sen1_1}-\eqref{eq_SenPCC_1}}
        \State{$\partial J_{C_m}/\partial \pmb{a_n} \leftarrow$ \eqref{eq_local_con}, \eqref{eq_Sen11_1}-\eqref{eq_Sen12_1}}    
        \State{$\partial \pmb{a_n}/\partial \pi_n \leftarrow$ \eqref{eq_step_2}}
        \State{$\partial \pi_n/\partial \pmb{\mu_n} \leftarrow$ \eqref{eq_step_31}}  
        \State{$\partial \pi_n/\partial \Sigma_n \leftarrow$ \eqref{eq_step_32}}   
        \State {$\partial \pmb{\mu_n}/\partial \pmb{\theta_{\mu_n}} \leftarrow$ $DNN_{\mu_n}$ [Back-propagation]}
        \State {$\partial \Sigma_n/\partial \pmb{\theta_{\Sigma_n}} \leftarrow$ $DNN_{\Sigma_n}$ [Back-propagation]}
        \State {$\pmb{g_{\mu_n}}, \pmb{b_{m,\mu_n}} \leftarrow$ \eqref{eq2_change_1} [Chain rule]}
        \State {$\pmb{g_{\Sigma_n}}, \pmb{b_{m,\Sigma_n}} \leftarrow$ \eqref{eq2_change_2} [Chain rule]}         
        \State {$H_n \leftarrow$ \eqref{eq2_H} [FIM Construction]}
        \State {Initialize $\pmb{\lambda_n}(k_0)$}        
        \For{$k \gets 1$ to $k^{max}$} 
        \State {$\pmb{\bar{\lambda}_n}(k) \leftarrow$ \eqref{eq2_DOpt_1} [Averaging operation]} 
        \State {$\pmb{\bar{\theta}_n}(k) \leftarrow$ \eqref{eq2_DOpt_2} [Primal gradient update]}        
        \State {$\pmb{\theta^t_n}(k+1) \leftarrow$ \eqref{eq2_DOpt_3}-\eqref{eq2_DOpt_5} [Projection on $M^L$]}
        \State {$\pmb{\lambda_n}(k+1) \leftarrow$ \eqref{eq2_DOpt_3_1} [Dual gradient update]} 
   \If {$||\pmb{\theta^t_{n}}(k+1)-\pmb{\theta^t_{n}}(k)|| \leq \Delta\theta_{n}$}
    \State $\pmb{\theta^{t+1}_{n}}\leftarrow\pmb{\theta^t_{n}}(k+1)$; Break;
    \EndIf
   \EndFor   
   \If {$||\pmb{\theta^{t+1}_{n}}-\pmb{\theta^{t}_{n}}|| \leq \Delta\theta_n$}
    \State Output $\pmb{\theta^{*}_{n}}\leftarrow\pmb{\theta^{t+1}_{n}}$; Break;
   \EndIf
   \EndFor
\State {Output well-trained parameterized policy $\pi_{n}(\pmb{\theta^{*}_{n}})$}   
\end{algorithmic}
\end{algorithm}

\subsection{Online Action Selection}\label{subsec:online}
The trained policy functions are used by the MG agents for online action selection. This process can be simply represented as sampling from the learned Gaussian policy functions \eqref{eq2_12_1}. First, the agents receive the latest values of the states, including the predicted solar irradiance and aggregate internal load power of MGs. These values are inserted into the trained DNNs \eqref{DNN_1} and \eqref{DNN_2} to obtain the mean and covariance matrices of the policy functions. Finally, samples are generated from the multivariate Gaussian distributions. These samples are averaged and passed to the local controllers of each controllable asset as a reference signal. 

\subsection{Backtracking Strategy}\label{subsec:backtracking}
Due to convex approximations in the formulations \eqref{eq2_7}-\eqref{eq2_9}, it is possible for few global constraints to be marginally violated in practice. To ensure feasibility, we can add a backtracking strategy into the proposed solution. This closed-loop backtracking strategy consists of two components, as shown in Fig. \ref{back_tracking}:
\begin{figure}
	\vspace{-0pt} 
	\vspace{-0pt}
	\centering
	\includegraphics[width=0.65\linewidth]{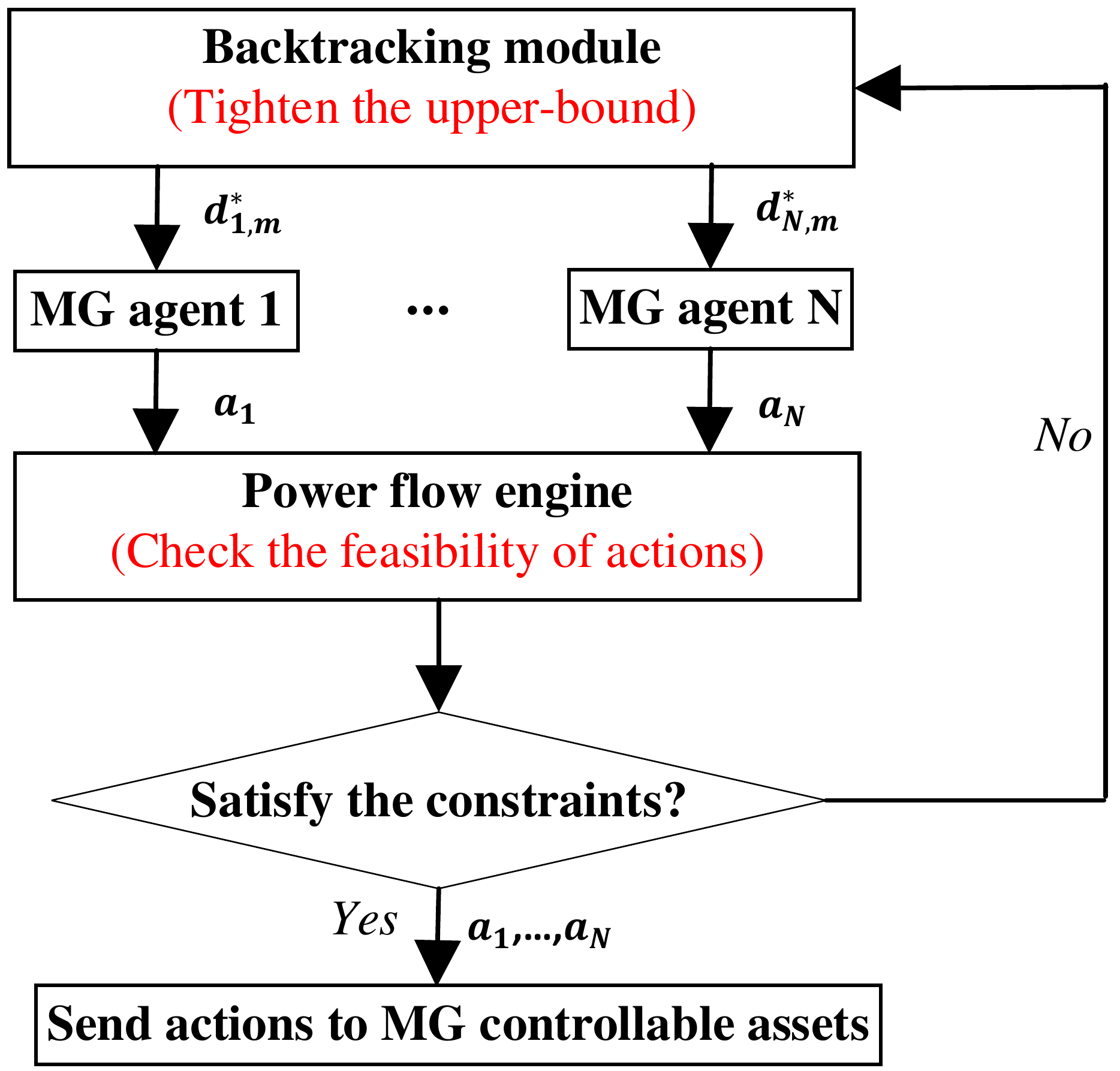}
	\vspace{-0pt} 
	\caption{Flowchart of the backtracking strategy}
	\centering
	\label{back_tracking}
    \vspace{-0pt} 
\end{figure} 

\textbf{Component 1.} \textit{Power flow engine (PFE):} The PFE receives the control actions from MG agents and runs a simple power flow program to obtain the status of all constraints. If no constraint is violated, the control signals are passed to controllable assets. If some constraints are violated, then the PFE will engage the backtracking process.

\textbf{Component 2.} \textit{Backtracking module:} The backtracking module tightens the upper-bound limit ($d_m$) (only) for the constraints that have been violated. The parameters of the trained DNNs will be re-updated according to update rules \eqref{eq2_DOpt_1}-\eqref{eq2_DOpt_3_1} and with the modified upper-bounds. The purpose of tightening the upper-bound is to provide a safety margin. In this paper the tightening process is performed using a user-defined coefficient multiplier, $0<\tau < 1$, as follows:
\begin{equation}
d_m^{*}=\tau d_m\label{eq_backtracking}
\end{equation}

\section{Simulation Results}\label{sec:Results}
\begin{figure}
\centering
\subfloat[33-bus system for distribution network \label{fig.5.0a}]{
\includegraphics[width=0.8\linewidth]{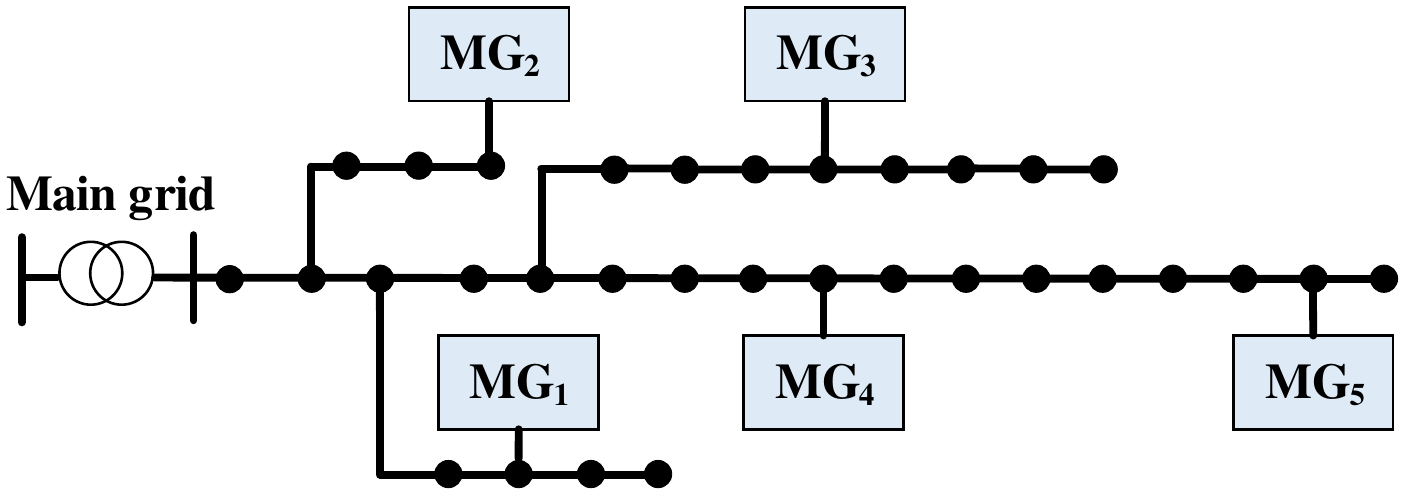}
}
\hfill
\subfloat[13-bus system for MGs \label{fig.5.0b}]{
\includegraphics[width=0.5\linewidth]{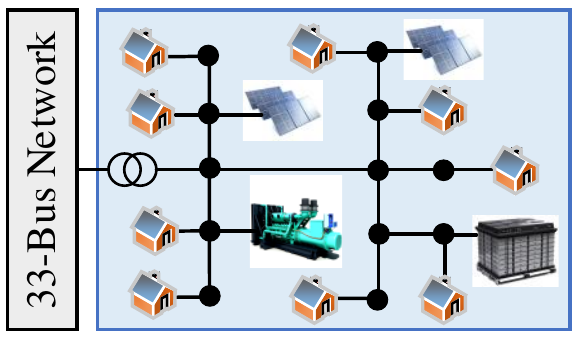}
}
\caption{Test system under study.}
\label{fig.5.0}
		\vspace{-0pt}\
\end{figure}

The proposed method is tested on a modified 33-bus distribution network \cite{13_bus}, which consists of five MGs as shown in Fig. \ref{fig.5.0a}. Each MG is modeled as a modified IEEE 13-bus network \cite{13_bus} at a low voltage level as shown in Fig. \ref{fig.5.0b}. When calculating the gradient factors, a single-phase AC power flow model is used for the sake of brevity. In the case study, the base power value is 100 kVA and base voltage values in the 33-bus distribution network and 13-bus MG networks are 12.66 kV and 4.16 kV, respectively.

The input data for load demands and PV generations have 15-minute time resolution are obtained from smart meter database \cite{Load_PV} to provide realistic numerical experiments. The assumption in this paper is that smart meters are installed throughout the network and the agents have access to a diverse data. The training and testing datasets are selected through uniform randomization to ensure that the proposed solver functions reasonably. Here, 1-month of the randomly selected data is used for testing and 11-month of the data is used for training. The energy price for the transferred power at the MG PCCs and the fuel price for the local DGs are adopted from \cite{EIA} and \cite{EIA_fuel}, respectively. The quadratic polynomial parameters of DG fuel consumption are adopted from \cite{Ali2015}. Table \ref{table_Parm} presents selected parameters for operational cost calculation in simulations. The average capacities for DGs in MGs are 60 kWh. The average capacities for ESSs in MGs are 20 kWh, the maximum charging/discharging rate is 4kW and the charging/discharging efficiencies are 95\% and 90\%, respectively.
\begin{table}[]
		\centering
		\renewcommand{\arraystretch}{1.3}		
		\caption{Selected cost function parameters}
	    \label{table_Parm}
\begin{tabular}{ccc}
\hline\hline
Description                                           & Notion            & Value            \\ \hline
Average electricity price ($\$/kWh$)                  & $\lambda^R$       & 0.046           \\ 
Average DG fuel price ($\$/L$)                        & $\lambda^f$       & 0.57              \\ 
Fuel cons. quadratic function parameter ($L/kW^2$)       & $a^f$          & 0.0001773        \\ 
Fuel cons. quadratic function parameter ($L/kW$)         & $b^f$          & 0.1709           \\ 
Fuel cons. quadratic function parameter ($L$)            & $c^f$          & 14.67          \\ \hline\hline
\end{tabular}
\end{table}

All the case studies are simulated using a PC with Intel Core i7-4790 3.6 GHz CPU and 16 GB RAM hardware. The simulations are performed in MATLAB \cite{Matlab} and OpenDSS\cite{Opendss} to obtain the gradient factors, update the learning parameters, solve the distributed training problem, and validate the results. In training, each episode is a learning update iteration based on the data that comes from one moving decision window. The length of the moving window is 4 samples with a 15-minute time step, which gives us a 1-hour window. The activation functions of each layer (including the output layer) of the feedforward networks are hyperbolic tangent-sigmoid (tansig). After various numerical tests, the parameters $\pmb{\theta_{\mu}}$ and $\pmb{\theta_{\Sigma}}$ of the neural networks are initialized using uniform distributions defined over the intervals (0,0.2) and (-0.03,0.03), respectively. In our simulations, we have observed that $\tau=0.9$ is sufficient for ensuring feasibility for those constraints that have been marginally violated after one-to-two rounds of backtracking. Table \ref{table_PL_Parm} summarizes selected DNN hyperparameters and other user-defined coefficients in simulations. The hyperparameters were optimized using a randomly-selected validation set (2 months worth of data) and Bayesian optimization with uninformative priors in MATLAB environment.  
\begin{table}[]
		\centering
		\renewcommand{\arraystretch}{1.3}		
		\caption{Selected DNN Hyperparameters and User-defined Coefficients}
	    \label{table_PL_Parm}
\begin{tabular}{ccc}
\hline\hline
Description                                 & Notion         & Value            \\ \hline
Length of the decision window in episode    & $T$            & 4                \\ 
Discount factor                             & $\gamma$       & 0.99             \\ 
Step size for updating $\theta$             & $\delta$       & $1\times10^{-3}$ \\ 
Maximum iteration                           & $k^{max}$      & 200              \\ 
Weight assigned to received information     & $w_{n}$        & 0.2              \\ 
Step size for primal gradient update            & $\rho_1$       & 0.01             \\ 
Step size for dual gradient update    & $\rho_2$       & 0.01             \\ 
Threshold for parameter updating            & $\Delta\theta$ & $1\times10^{-4}$ \\ 
Tightening multiplier                       & $\tau$         & 0.9              \\ 
Number of hidden layer                      & -              & 3                \\
Number of neurons per hidden layer          & -              & 10               \\
Size of minibatches                         & -              & 128              \\ 
Activation function of DNNs      & -              & \textit{tansig}  
\\ \hline\hline
\end{tabular}
\end{table}

Further, to demonstrate the effectiveness of SMAS-PL, three benchmark methods have been considered, including an optimization-based method, an on-policy method and an off-policy method. The first benchmark method is an optimization-based method, which leverages YALMIP toolbox to solve the optimal power management of networked MGs using IBM ILOF CPLEX 12.9. The second one is the unconstrained policy gradient learning (U-PL) method, which leverages the same algorithm as the proposed SMAS-PL, however, certain constraints are removed during the training process of U-PL. By comparing the SMAS-PL and the U-PL, we can show the effectiveness of the SMAS-PL when handling different local and global constraints. The U-PL can be considered as an on-policy benchmark. We also consider an off-policy benchmark method, namely the deep Q-network (DQN). In \cite{Constrined_EV,DQN}, DQN uses deep neural networks (DNNs) to approximate the Q-function and provide Q-value estimation for discretized control actions. To include the constraints in DQN, we have followed the suggestion in \cite{Constrined_EV,penalty} and added penalty terms to the reward function of the benchmark DQN to discourage constraint violation. The penalty coefficients for global and local constraints are manually tuned based on the DQN performance. However, since the benchmark DQN was not originally designed for continuous actions, we have first discretized the agents' action space with a step size of 33\% of the constraint upper limit. For example, if the upper limit of a diesel generation (DG) power output is 60 kW, then, the power output action of DG has been discretized as (0,20,40,60) kW. Similar discretization has been applied to the actions of PV inverters and ESSs.The inputs of the DNN are the system states, and the outputs of the DNN are estimations for the Q-value function for each discrete action. The DNN is parameterized as a function approximator to represent the Q-value function. The temporal difference (TD) learning algorithm is used to train the DNN by minimizing the mean-squared TD error. The discount factor and learning rate in DQN are set to the same values as those of SMAS-PL. The exploration factor is set to 0.1 in the $\epsilon$-greedy action selection of DQN. The structure of DNN in the benchmark DQN has been obtained using cross-validation. The dimensions of the input and output layers have been extended by the number of MG agents and the number of discrete actions. Note that the benchmark U-PL is implemented in a multi-agent framework, while the benchmark DQN is implemented in a centralized way.

\subsection{System Operation Outcomes}
\begin{figure} [h]
	\vspace{-0pt} 
	\vspace{-0pt}
	\centering
	\includegraphics[width=0.8\linewidth]{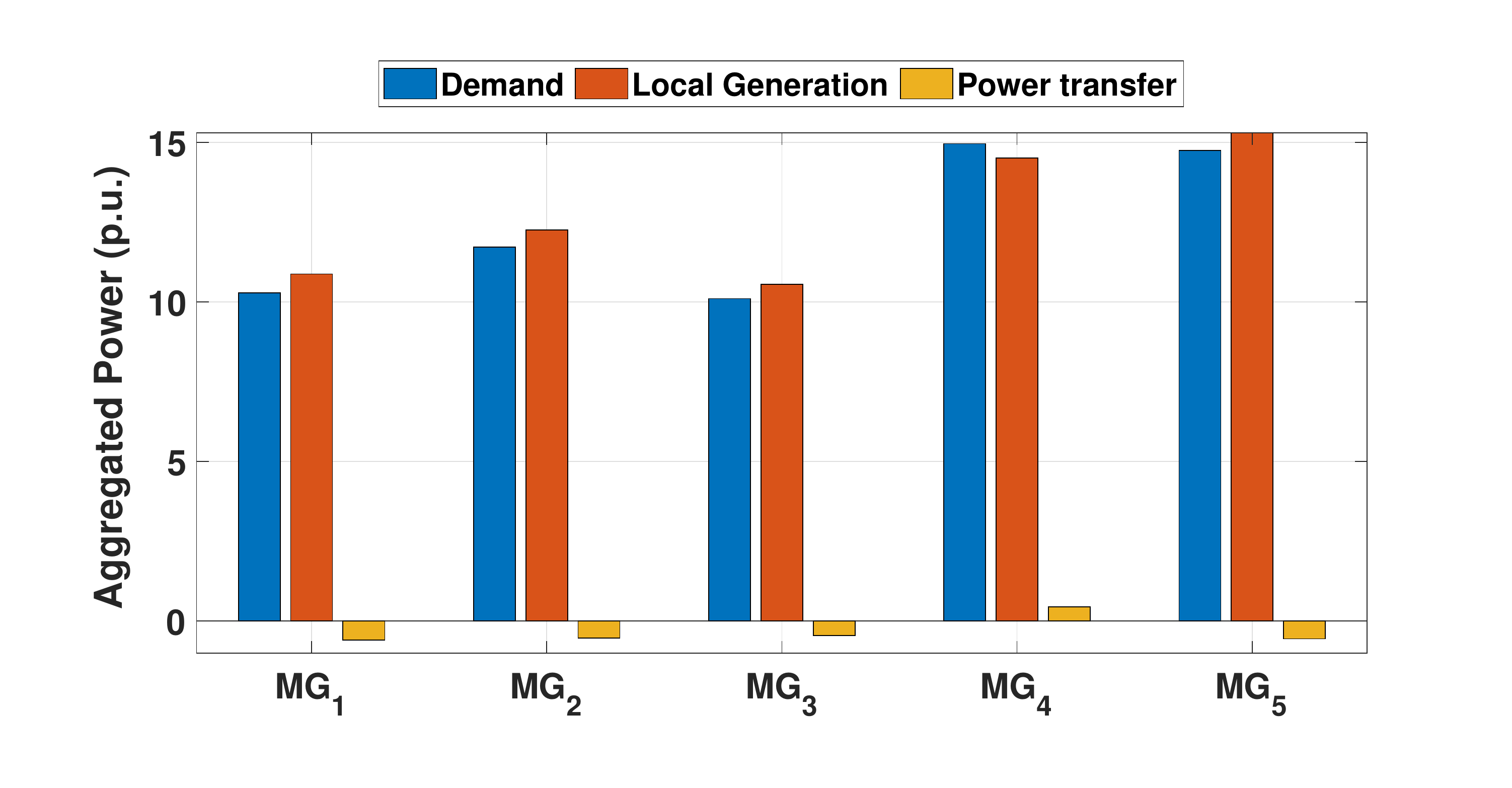}
	\vspace{-0pt}\
	\caption{Aggregated power of local demand, local generation and power transfer for $MG_1$-$MG_5$.}  
	\centering
	\label{fig.5.1}
	\vspace{-0pt}
\end{figure}

\begin{figure} [h]
	\vspace{-0pt} 
	\vspace{-0pt}
	\centering
	\includegraphics[width=0.8\linewidth]{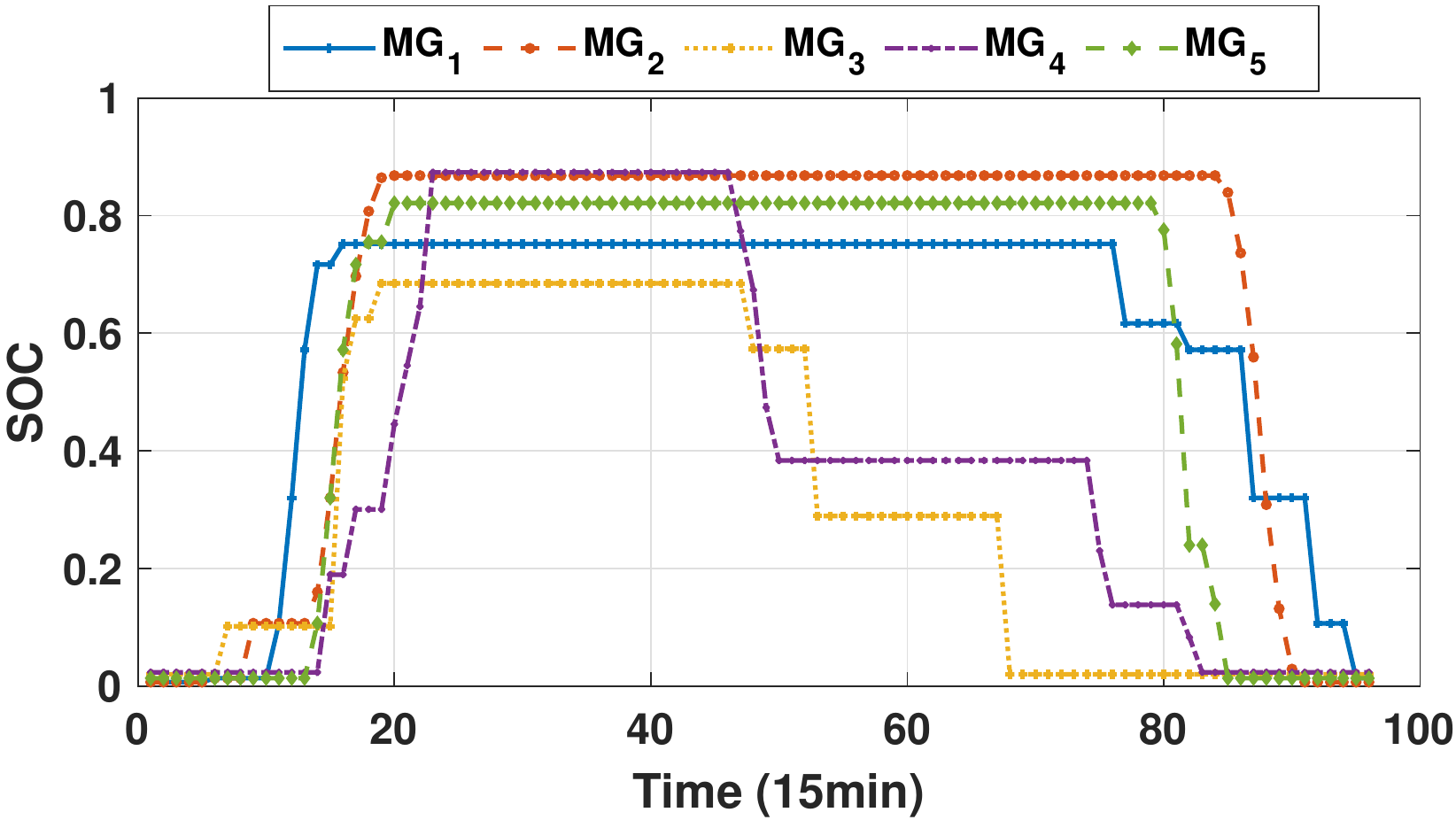}
	\vspace{-10pt}\
	\caption{ESS dispatching results for $MG_1$-$MG_5$.}
	\centering
	\label{fig.5.2}
	\vspace{-0pt}
\end{figure}

In the case study, action selection is performed by sampling 100 times from the trained policy functions (distributions). Then the dispatching action is obtained by averaging the selected samples. A trade-off is involved in choosing the number of action samples: if this number is too large, then the selected actions will converge to the policy mean, which implies that model uncertainties are ignored. This could result in erroneous and sub-optimal solutions in case the learned model is over-fitting (i.e., when the estimated mean has large errors). On the other hand, if the number of samples is too small, then the outcomes can deviate from the learned mean value, which can also result in low-quality outcomes. The average outcomes are shown in Fig. \ref{fig.5.1}, Fig. \ref{fig.5.2} and Table \ref{table_Com}. The aggregate MG demand, aggregate MG generation, and aggregate power transfer through PCCs of MGs over a day are shown in Fig. \ref{fig.5.1}. It can be seen that the main MG demands are supplied by the local generation within MGs due to low DG fuel prices and renewable outputs. While most MGs are exporting power to the upstream distribution grid, $MG_4$ is importing power to satisfy the heavy local load that cannot be fully supplied internally. In all cases the power balance is maintained within the MGs. The ESS SOCs for each MG are shown in Fig. \ref{fig.5.2}, where can be seen that ESSs charge during off-peak period and discharge during peak time to provide optimal power balancing support for MGs. Table \ref{table_Com} presents comparisons between the benchmark optimization-based method, the benchmark DQN and the proposed SMAS-PL, including the average daily cost of operation over numerous scenarios, average online decision time, and MG privacy maintenance.   

\begin{table}[]
		\centering
		\renewcommand{\arraystretch}{1.3}		
		\caption{Comparison between Centralized Solver, DQN and SMAS-PL Method}
		\label{table_Com}
			\vspace{-0pt}\
\begin{tabular}{cccccc}
\hline\hline
                       & Cen. solver & DQN & SMAS-PL   \\ \hline
Average daily cost (\$)        & 1356.60 & 1928.4 & 1372.11  \\ \hline
Average time (second)       & 145.50 & 10.30 & 1.40 (per agent)   \\ \hline
MG privacy maintenance & No   & No   &  Yes  \\ \hline\hline
\end{tabular}
\end{table}

In general, the SMAS-PL method has three fundamental advantages over centralized optimization method: 1) Even though the offline training process in our method takes a long time (around 35 minutes per agent), the average online decision time for the proposed SMAS-PL is about only 1.4 seconds per agent, which is much shorter than the average time 145.5 seconds for the centralized optimization solver. Thus, the real-time response of the trained policy function is almost 100 times faster than that of the OPF solver. The reason for this is that the OPF solver needs to find the optimal solution of a complex optimization problem in real-time, while our approach simply samples from multivariate Gaussian distributions that embody optimal control policies. Furthermore, we have observed that the computational cost of the centralized OPF solver rises almost quadratically with the size of the system; beyond a certain point the commercial solver is not able to provide solutions in a reasonable time. On the other hand, our SMAS-PL retains an almost constant online decision time, while the cost of offline training increases almost linearly. 2) The proposed PL method takes advantage of a multi-agent (distributed) framework to train the policy function of each MG agent; in practice, this distributed framework can be implemented using parallel computation techniques, which also enhances the scalability of the proposed SMAS-PL method compared to centralized solvers. 3) Due to its distributed nature, the proposed SMAS-PL method maintains the privacy and data ownership boundaries of individual MGs. During the training process, the MG agents do not need to share control policy parameters, policy functions, cost functions, and local asset constraints with each other. The only variables that are shared among MG agents are the Lagrangian multipliers corresponding to global network constraints. These multipliers do not have a physical meaning and thus, do not contain sensitive information.  

Based on the comparison between the centralized solver and our proposed method, there is still a 1.14\% difference between the solutions from  the centralized solver and the SMAS-PL method, which might be caused by the following reasons: (i) Unlike the centralized solver, which has access to the full systemic model information, and thus, can guarantee at least a local optimal solution, the proposed SMAS-PL method lacks a guarantee of optimality. Also, in order to obtain a high-quality solution, the SMAS-PL needs to first approximate the original problem with a convex surrogate, which despite enhancing the problem tractability, comes at the expense of loss of accuracy and a reduction in performance. (ii) The proposed backtracking mechanism is a heuristic strategy, which is aimed at obtaining a feasible solution that might come at the expense of a loss in the reward. (iii) To obtain a consensus-based solution, the SMAS-PL needs a reliable inter-agent communication infrastructure, which could be costly. (iv) In case of changes in system structure, the SMAS-PL will need an offline re-training phase to adapt to new system conditions. This could take some time, during which the agents will experience a temporary decline in their payoffs. The comparison between DQN and SMAS-PL is discussed in Section \ref{sec:Algorithm Performance}. 

\subsection{Algorithm Performance}\label{sec:Algorithm Performance}
\begin{figure}
\centering
\subfloat[Selected $\theta_{\mu}$'s during training process\label{fig.6.1a}]{
\includegraphics[width=0.8\linewidth]{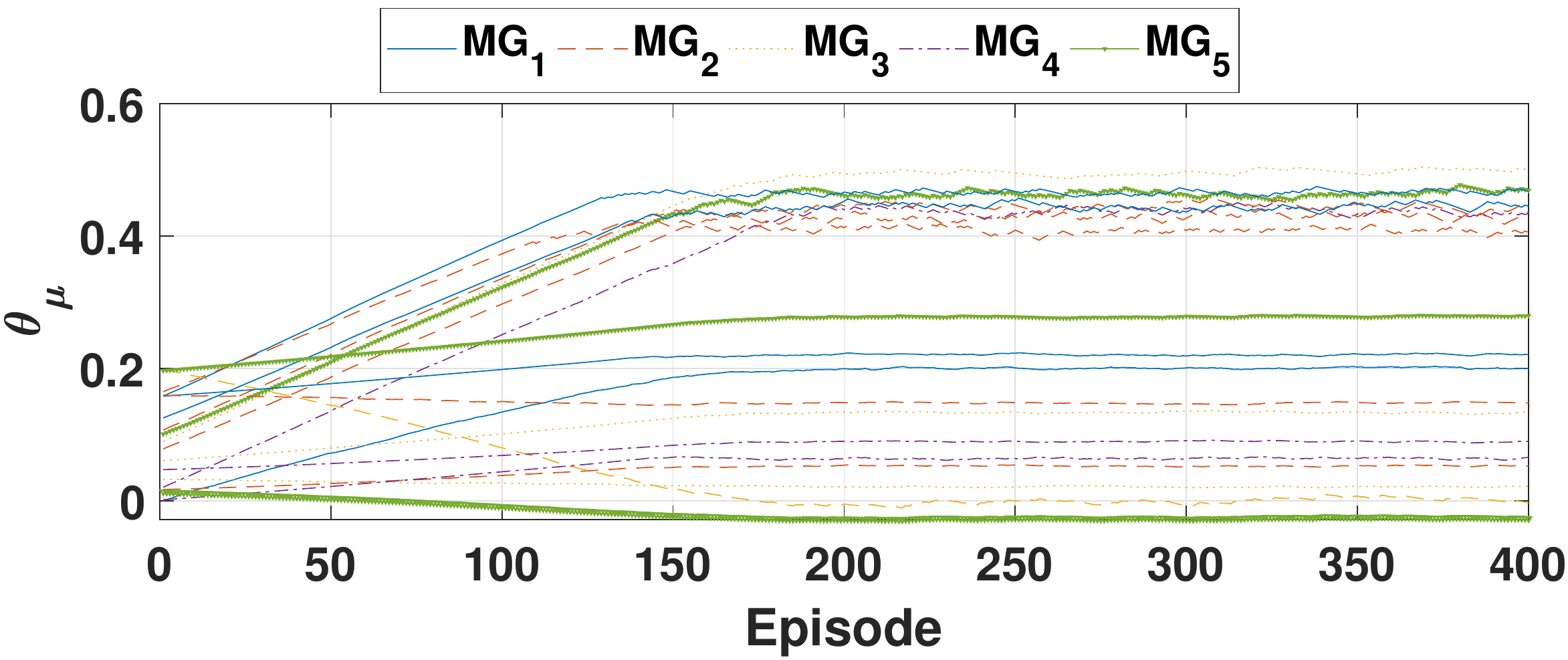}
}
\hfill
\subfloat[Selected $\theta_{\Sigma}$'s during training process\label{fig.6.1b}]{
\includegraphics[width=0.8\linewidth]{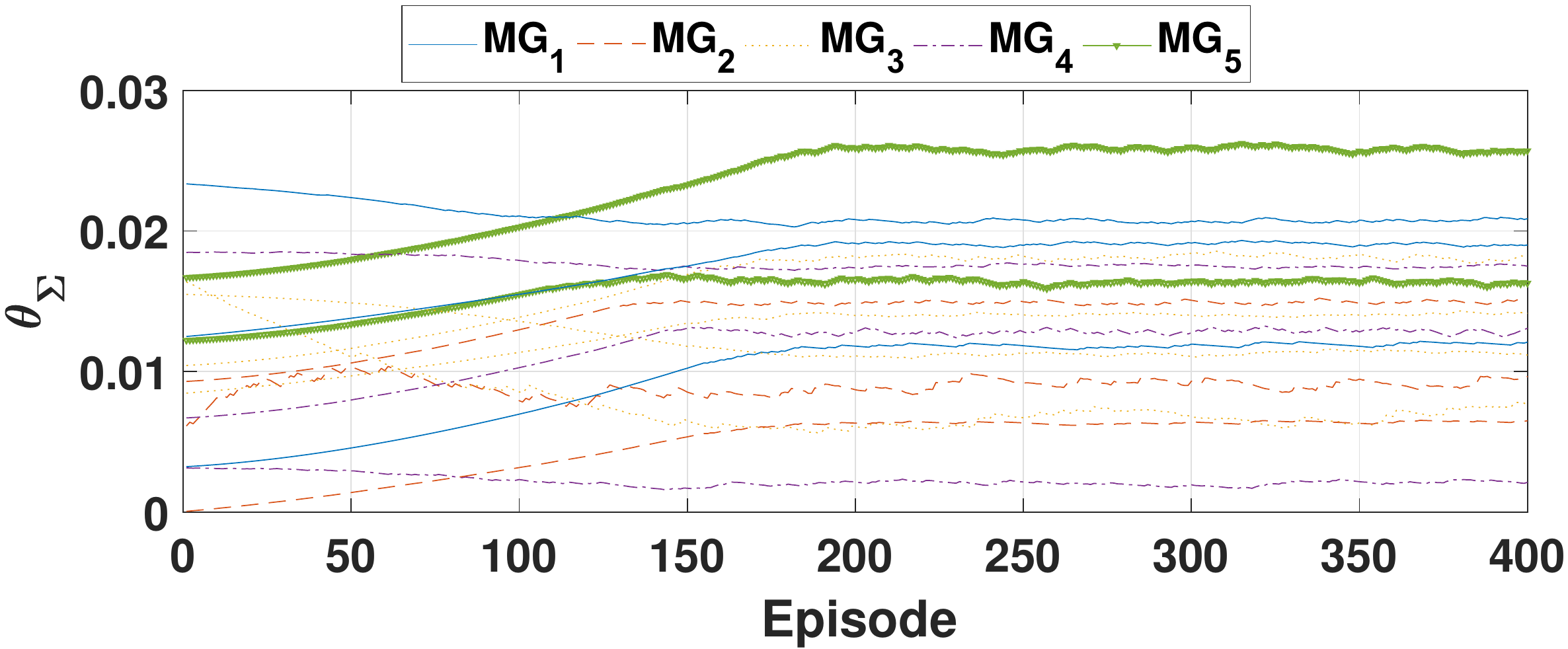}
}
\caption{Convergence of learning parameters $\theta_{\mu}$ and $\theta_{\Sigma}$ for $MG_1$-$MG_5$ .}
\label{fig.6.1}
		\vspace{-10pt}\
\end{figure}

Fig. \ref{fig.6.1a} and Fig. \ref{fig.6.1b} show the convergence of a selected group of learning parameters, $\pmb{\theta_{\mu}}$ and $\pmb{\theta_{\Sigma}}$ during the training process, for each MG agent. As can be seen, the changes in $\pmb{\theta_{\mu}}$ are relatively larger than that of $\pmb{\theta_{\Sigma}}$. This is due to the higher levels of sensitivity of MG agents' objective functions to the mean values of the control actions compared with their variance levels. 

\begin{figure} [h]
	\vspace{-0pt} 
	\vspace{-0pt}
	\centering
	\includegraphics[width=0.8\linewidth]{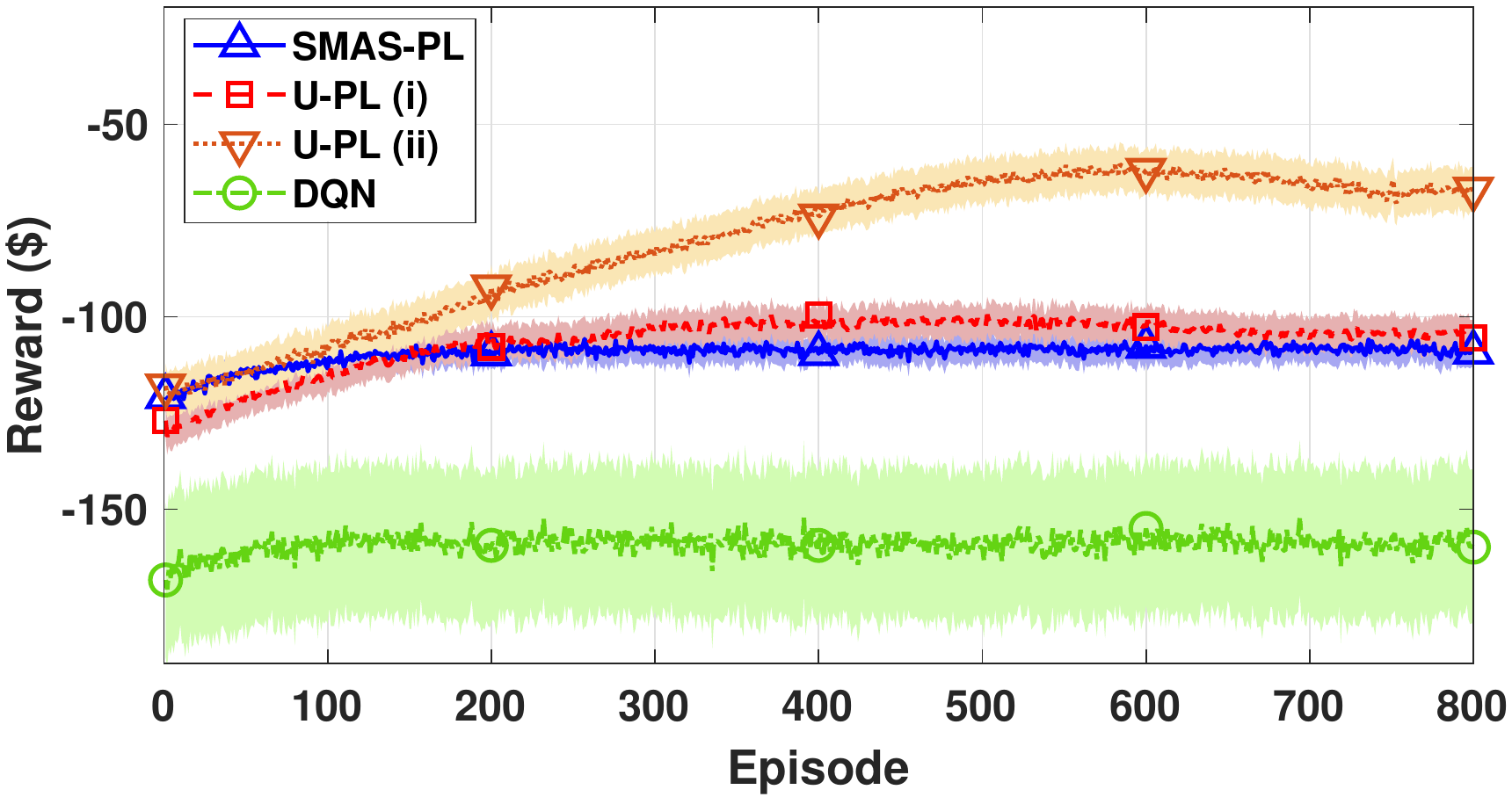}
	\vspace{-10pt}\
	\caption{Comparison of the average hourly rewards with different methods.}
	\centering
	\label{fig.6.2}
	\vspace{-5pt}
\end{figure}

In Fig. \ref{fig.6.2}, the average hourly rewards under SMAS-PL, U-PL, and DQN are compared with each other. Note that here, the moving average rewards and the episodic rewards of different methods are depicted by dark and light curves. It can be observed that SMAS-PL and U-PL both outperform DQN in term of the total reward. The reason for this is that the SMAS-PL and U-PL leverage the proposed iterative and distributed technique to adaptively tune the Lagrangian multipliers through information exchange between MG agents; on the other hand, the DQN needs to manually design penalty coefficients for constraint violations, which either offers inadequate penalization of the constraint violations or excessive punishment for the constraints. Also, SMAS-PL and U-PL have continuous action spaces, while DQN employs action discretization, which hinders accurate exploration of action space. After the NNs are fully trained, the SMAS-PL samples the actions from the learned multivariate Gaussian distributions that embody optimal control policies, while the benchmark DQN selects the control actions that have the highest estimated Q-values for the given state according to the trained DNN. Based on the results in Table \ref{table_Com}, the decision time for the SMAS-PL is around 1.4 seconds per agent, while the decision time for the benchmark DQN is approximately 10.3 seconds. Thus, the decision time for the proposed SMAS-PL is faster than the benchmark DQN, because the multi-agent framework enables the SMAS-PL to sample decision actions in parallel for each MG agent, while the benchmark DQN selects the control actions for all the MGs together in a centralized way. Two cases are considered in implementing U-PL: (i) no DG capacity constraints for $MG_1$ and $MG_2$; (ii) no DG capacity constraints for $MG_1$-$MG_5$. In cases (i) and (ii) of the U-PL, the agents obtain a higher reward compared to the SMAS-PL due to the constraint omission; however, this comes at the expense of decision infeasibility. In case of the SMAS-PL, these operational constraints are satisfied, which also leads to a drop in total reward, as expected. This shows that our proposed constrained PL decision model can ensure the feasibility of the control actions w.r.t. the constraints of the power management problem. Note that even though the proposed SMAS-PL framework is similar to the TRPO \cite{ConPolOpt2017}, the TRPO has theoretical guarantees for monotonic increase in return, while such guarantees do not exist for the approximate QCLP formulation in the proposed SMAS-PL. However, compared to TRPO our solution offers a simpler, more efficient, and tractable alternative, with fewer learning parameters.

\begin{figure} [h]
	\vspace{-0pt} 
	\vspace{-0pt}
	\centering
	\includegraphics[width=0.8\linewidth]{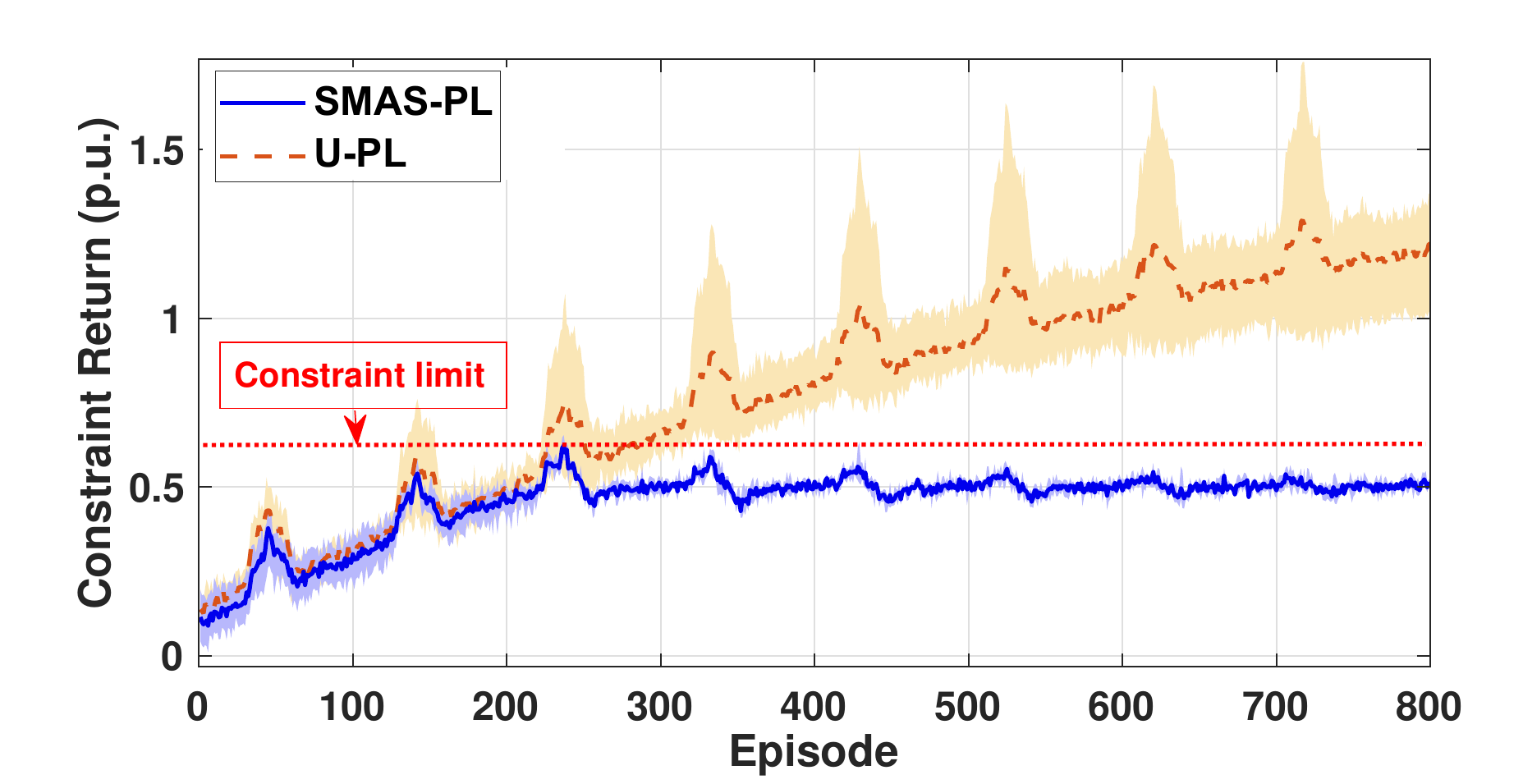}
	\vspace{-10pt}\
	\caption{Comparison of constraint values w/ and w/o DG capacity constraints in $MG_1$.}  
	\centering
	\label{fig.6.3}
	\vspace{-0pt}
\end{figure}

Furthermore, Fig. \ref{fig.6.3} shows the constraint values during the training iterations for a 1-hour time window, for the two cases with and without DG capacity constraints in $MG_1$, where the dark blue and red curves represent averaged constraint values, and the light blue and red areas represent the variations around the average curves for the SMAS-PL and U-PL, respectively. During the training process, the U-PL violates the upper boundary for DG generation limit (i.e., local constraint case study); on the other hand, the SMAS-PL solver satisfies the DG generation capacity constraints, which implies that the local constraints can be safely maintained. Therefore, compared to U-PL, the proposed SMAS-PL has shown to be able to generate control actions that not only improve the reward function but also satisfy the constraints.

\begin{figure} [h]
	\vspace{-0pt} 
	\vspace{-0pt}
	\centering
	\includegraphics[width=0.8\linewidth]{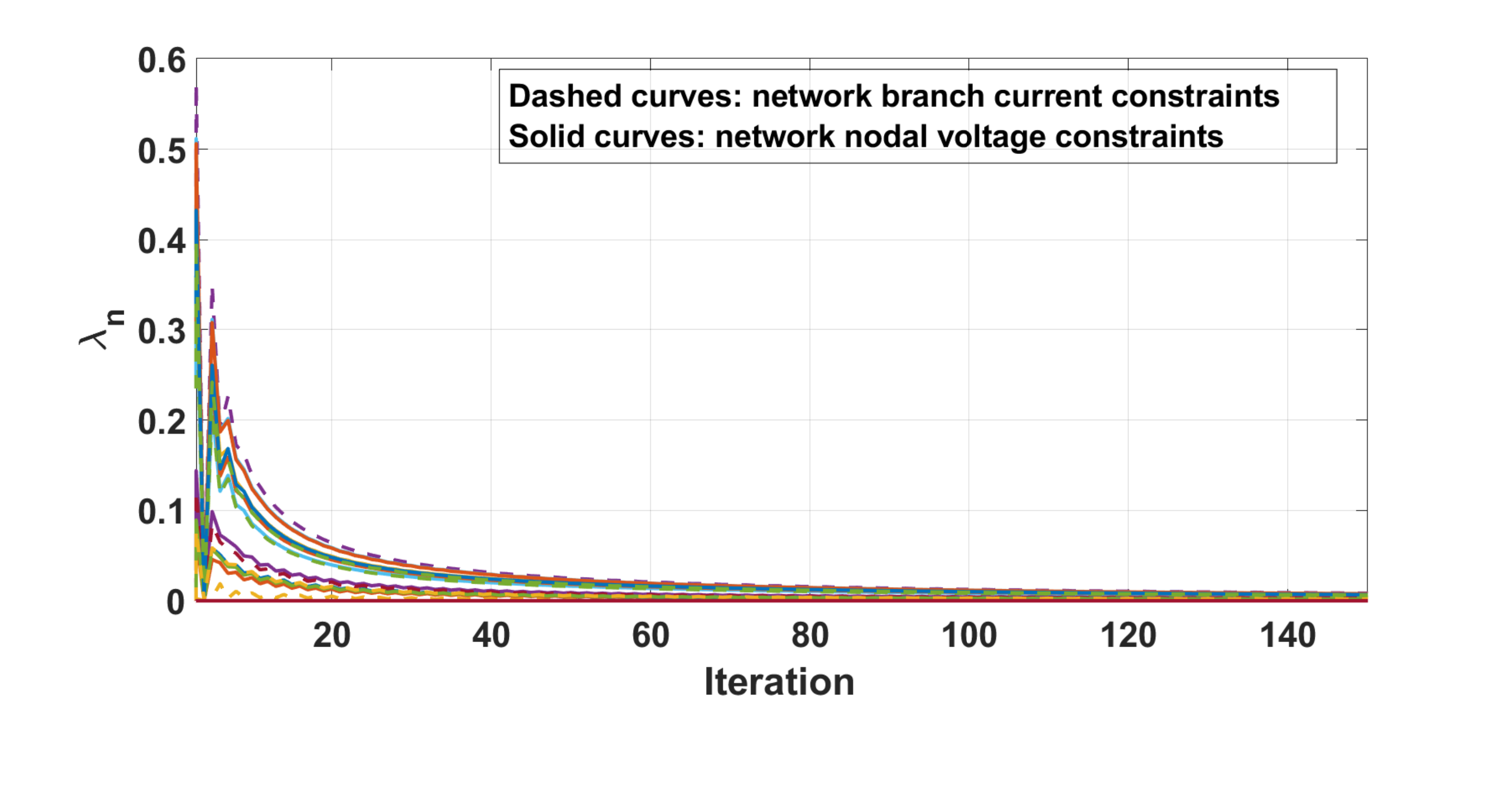}
	\vspace{-10pt}\
	\caption{The performance of the iterative distributed training method in one episode (no binding global constraints).}
	\centering
	\label{fig.6.4}
	\vspace{-0pt}
\end{figure}

One example of the distributed training convergence process is shown in Fig. \ref{fig.6.4} for a policy gradient update step. As can be seen, the Lagrangian multipliers $\lambda_n$ reach zero over iterations of the proposed multi-agent algorithm, which indicates that all the global constraints, including nodal voltage and branch current limits, are satisfied and feasible solutions are obtained. This also means that the bus voltage and line current constraints are not binding for this case.

\begin{figure} [h]
	\vspace{-0pt} 
	\vspace{-0pt}
	\centering
	\includegraphics[width=0.8\linewidth]{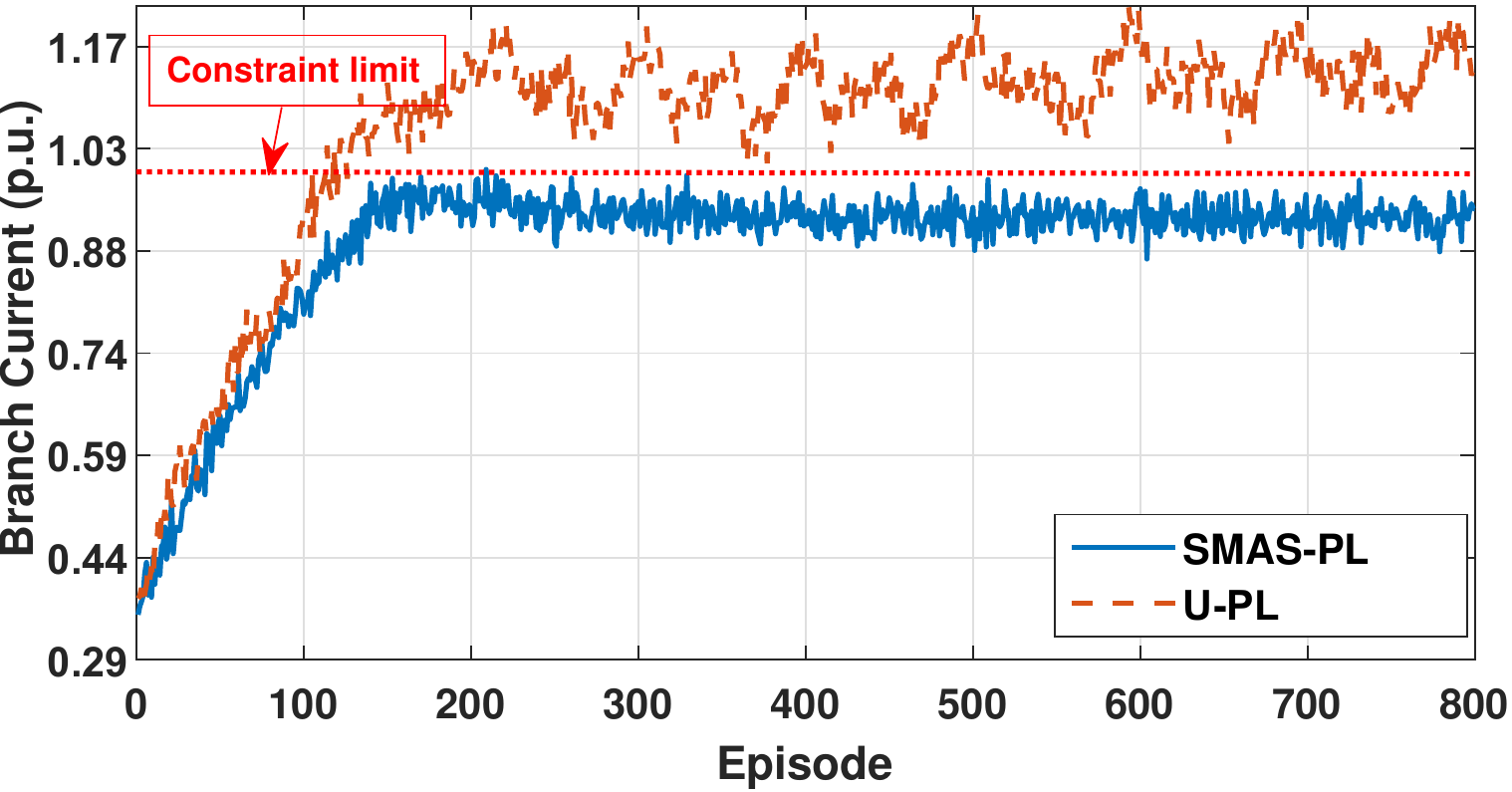}
	\vspace{-10pt}\
	\caption{Selected global branch current constraint return values for MG agents.} 
	\centering
	\label{fig.6.6}
	\vspace{-0pt}
\end{figure}

\begin{figure} [h]
	\vspace{-0pt} 
	\vspace{-0pt}
	\centering
	\includegraphics[width=0.8\linewidth]{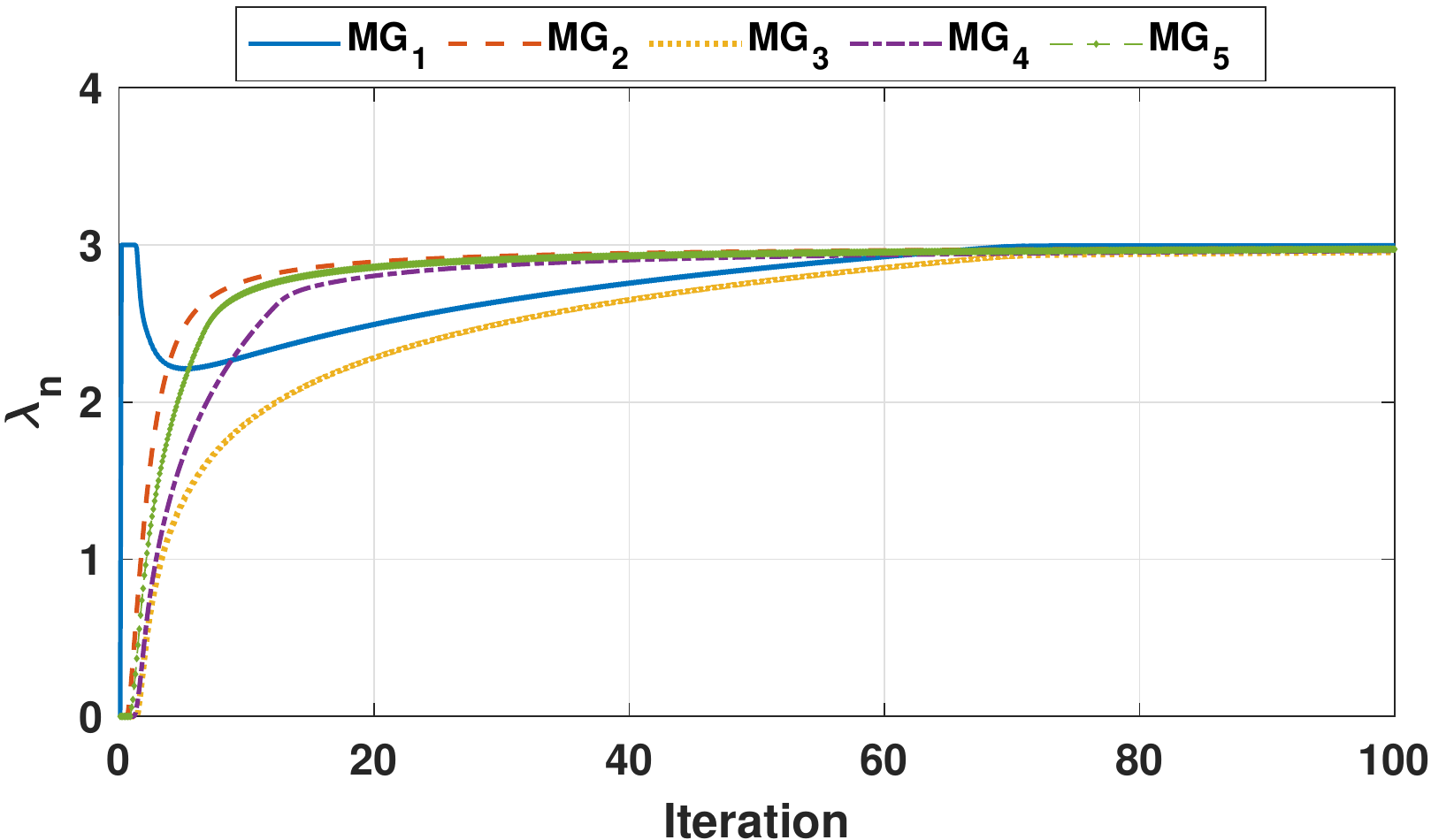}
	\vspace{-10pt}\
	\caption{MG agents' consensus on $\lambda_n$ for the selected global constraint.}  
	\centering
	\label{fig.6.5}
	\vspace{-0pt}
\end{figure}

Another example is given to demonstrate the effectiveness of the SMAS-PL in handling binding global constraints. This case shows a line flow constraint in the grid under the proposed SMAS-PL and a U-PL baseline; as is observed in Fig. \ref{fig.6.6}, the U-PL has generated infeasible decisions that violate the constraint, while our approach has prevented the flow to go above its upper bound. Further, as can be seen in Fig. \ref{fig.6.5}, the Lagrangian multipliers for this binding constraint reach a non-zero constant number over iterations. This also shows the agents' estimations of Lagrange multipliers for a global line flow constraint; as can be seen, using the proposed SMAS-PL the agents are capable of reaching consensus on the value of the multiplier without having any access to each other's policy functions, which corroborates the performance of our proposed method under incomplete information.

\begin{figure} [h]
	\vspace{-0pt} 
	\vspace{-0pt}
	\centering
	\includegraphics[width=0.8\linewidth]{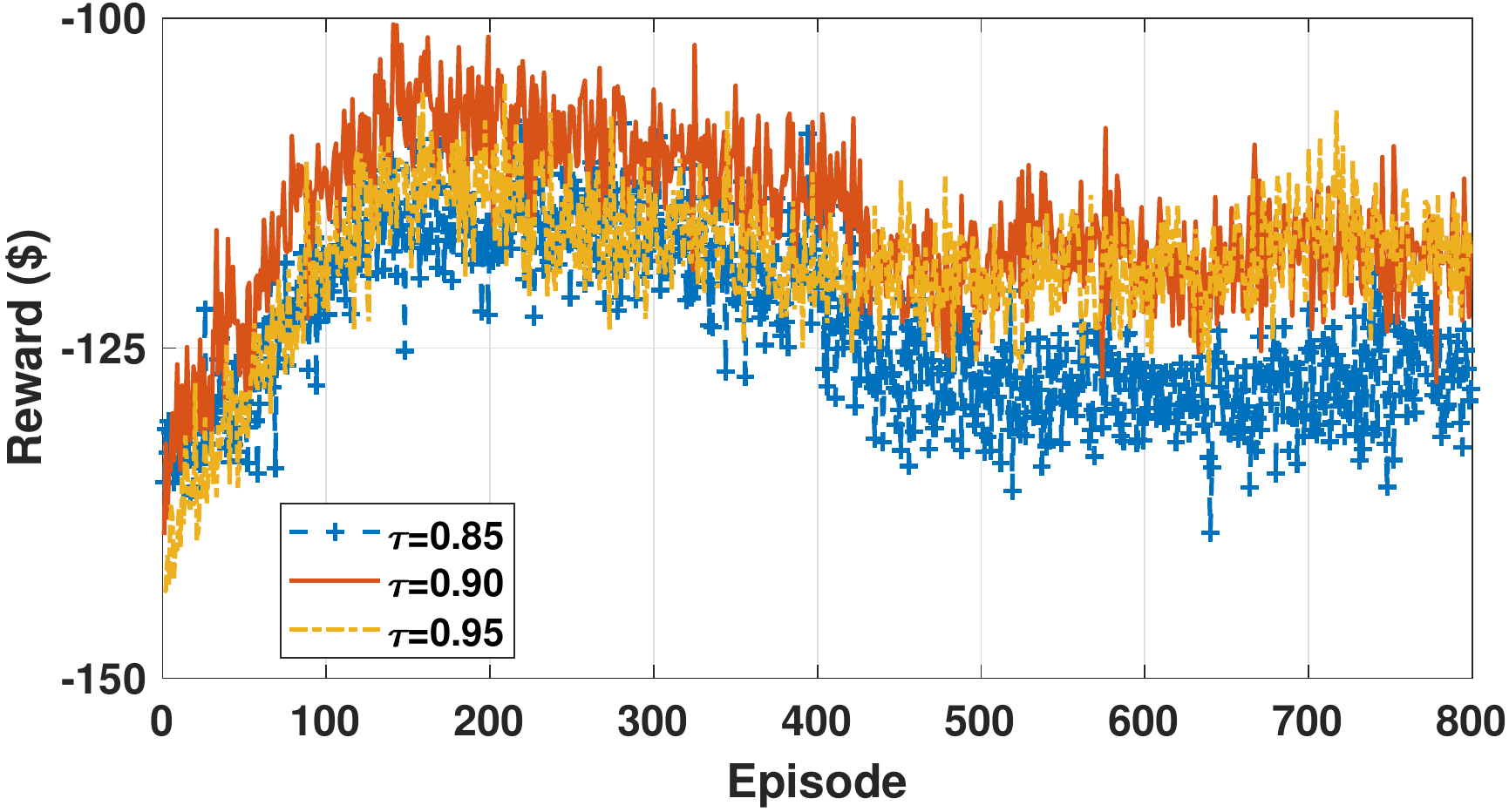}
	\vspace{-10pt}\
	\caption{Impact of backtracking on algorithm performance.}  
	\centering
	\label{fig.R_tau}
	\vspace{-0pt}
\end{figure}

To validate the tightening parameter levels ($\tau$), we have studied the impact of different $\tau$ values on the reward. Here, at episode 400, the value of $\tau$ is decreased from 1 to (0.95, 0.9, 0.85). The average rewards for different drops in $\tau$ are compared in Fig. \ref{fig.R_tau}. It can be observed that for values of $\tau$ close to 1 (i.e., $\tau$=0.95 and $\tau$=0.9) the reward values are very close to each other. However, as $\tau$ deviates from unity and reaches $\tau$=0.85, the reward drops significantly. In our simulation, we have observed that $\tau$=0.9 is sufficient for ensuring feasibility for those few constraints that have been marginally violated in certain operation scenarios after one-to-two rounds of backtracking. Note that this threshold needs to be fine-tuned for specific grids.

\begin{figure} [h]
	\vspace{-0pt} 
	\vspace{-0pt}
	\centering
	\includegraphics[width=0.8\linewidth]{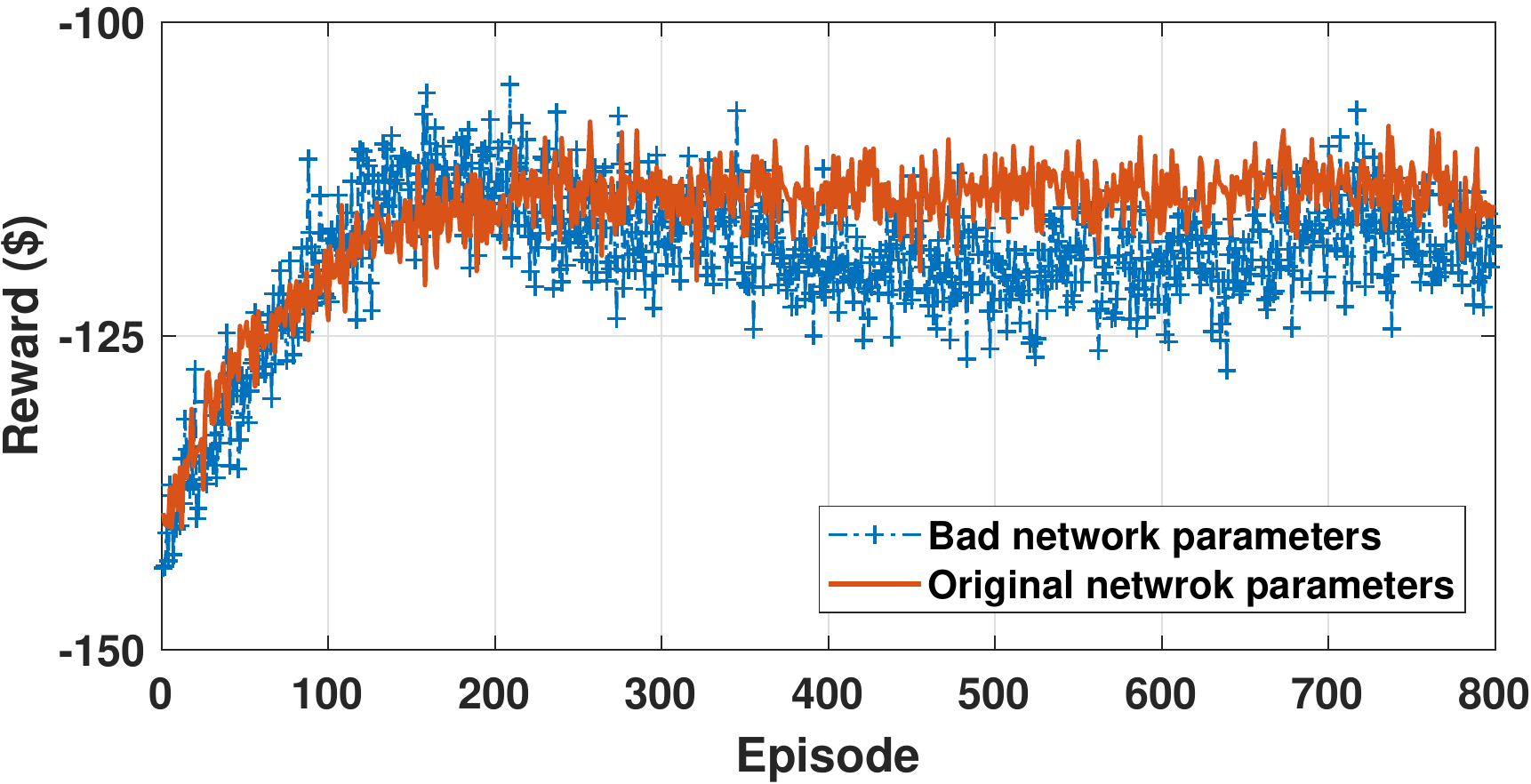}
	\vspace{-10pt}\
	\caption{Analysing the impact of bad network data on decision model outcomes.}
	\centering
	\label{fig.R_bad}
	\vspace{-0pt}
\end{figure}

To simulate the impact of bad network parameter data on training model performance, we have added random errors (with a 10\% variance) to the network resistance (R) and reactance (X) parameters during the training process. The bad network data will lead to errors in gradient factors \eqref{eqPF_8_1}-\eqref{eq_local_con} (see Appendix \ref{app:gradient}). To validate the SMAS-PL under network data imperfection, we have compared the average reward obtained with perfect knowledge of network parameters and under bad network parameter information. It can be observed in Fig. \ref{fig.R_bad}, even though the learning process with bad network data shows more volatility and needs more time to reach convergence, the model still reaches reward values close to the ideal case. However, due to the information imperfection, a loss of reward is inevitable.

\section{Conclusion}\label{sec:Con}
Conventional model-based optimization methods suffer from high computational costs when solving large-scale multi-MG power management problems. On the other hand, the conventional model-free methods are black-box tools, which may fail to satisfy the operational constraints. Motivated by these challenges, in this paper, a SMAS-PL method has been proposed for power management of networked MGs. Our proposed method exploits the gradients of the decision problem to learn control policies that achieve both optimality and feasibility. Furthermore, to enhance computational efficiency and maintain the policy privacy of the control agents, a distributed consensus-based training process is implemented to update the agents' policy functions over time using local communication.

Note that the current case study has been conducted over a balanced single-phase distribution system. However, our proposed SMAS-PL is not limited to single-phase distribution systems and can be potentially extended to unbalanced three-phase systems. One solution to this challenge could be using a single policy function for the resources connected to all phases (note that theoretically-speaking our method is not limited by the number of phases). However, this brute-force solution may lack scalability. Another solution extension to a multi-phase system cannot be fully addressed by having three separate policy functions per phase. A more efficient and scalable extension to unbalanced systems remains the subject of our future research.
 
\appendices
\section{Calculation of $\partial J_{R_n}/\partial \pmb{a_n}$ and $\partial J_{C_m}/\partial \pmb{a_n}$}\label{app:gradient}
The major difficulty in determining $\partial J_{R_n}/\partial \pmb{a_n}$ and $\partial J_{C_m}/\partial \pmb{a_n}$ pertains to the agents' reward functions and global constraint returns, \eqref{eq3_1}-\eqref{eqPF_i}, which are only implicitly related to the control actions. Since the reward and all the global constraint returns are functions of the observation variables, $\pmb{V}$ and $\pmb{I}$, the gradients of these variables w.r.t. control actions are obtained and used to quantify $\partial J_{R_n}/\partial \pmb{a_n}$ and $\partial J_{C_m}/\partial \pmb{a_n}$. To do this, a four-step process is proposed that leverages the current injection-based AC power flow equations:

\textbf{Step 1 - } First, the gradients of real and imaginary parts of nodal current injection w.r.t. control actions are derived (denoted as $\partial \pmb{I^{Re}}/\partial \pmb{a_n}$ and $\partial \pmb{I^{Im}}/\partial \pmb{a_n}$, respectively.) To achieve this, the nodal power balance and nodal current injection relationships in the network are employed \cite{Sen_2016}:
\begin{equation}
I^{Re}_{i,t'} = \frac{p_{i,n,t'}V^{Re}_{i,t'}+ q_{i,n,t'}V^{Im}_{i,t'}}{V_{i,t'}^{2}}\label{eqPF_8_1}
\end{equation}
\begin{equation}
I^{Im}_{i,t'} = \frac{p_{i,n,t'}V^{Im}_{i,t'}- q_{i,n,t'}V^{Re}_{i,t'}}{V_{i,t'}^{2}}\label{eqPF_8_2}
\end{equation}
\begin{equation}
p_{i,n,t'}= P_{i,n,t'}^{D}-P_{i,n,t'}^{DG}-P_{i,n,t'}^{PV}+P^{Ch}_{i,n,t'}-P^{Dis}_{i,n,t'}\label{eqPF_9}
\end{equation}
\begin{equation}
q_{i,n,t'}= Q_{i,n,t'}^{D}-Q_{i,n,t'}^{DG}-Q_{i,n,t'}^{PV}+Q^{ESS}_{i,n,t'}\label{eqPF_12}
\end{equation}
where, $I^{Re}_{i},I^{Im}_{i}$ and $V^{Re}_{i},V^{Im}_{i}$ denote the real and imaginary parts of nodal voltage and current injection at node $i$. Using these equations, $\partial \pmb{I^{Re}}/\partial \pmb{a_n}$ and $\partial \pmb{I^{Im}}/\partial \pmb{a_n}$ are derived and shown in Table \ref{table_Sen}. Note that the entries of this table can be calculated using the real and imaginary parts of nodal voltages, which in practice are either measured or estimated \cite{Sen_2016}. 
\begin{table}[]
		\centering
		\renewcommand{\arraystretch}{1.3}		
		\caption{Partial derivations of $\pmb{I^{Re}}$ and $\pmb{I^{Im}}$ w.r.t. $\pmb{a_n}=[P^{DG}_n,P^{Ch}_n,P^{Dis}_n,Q^{DG}_n,Q^{PV}_n,Q^{ESS}_n]$}
		\label{table_Sen}
			\vspace{-0pt}\
\begin{tabular}{l|ccccccc}
\hline\hline
\diagbox{-}{$\pmb{a_n}$}                        & $P^{DG}_{n,t'}$              & $P^{Ch}_{n,t'}$            & $P^{Dis}_{n,t'}$             & $Q^{DG}_{n,t'}$              & $Q^{PV}_{n,t'}$              & $Q^{ESS}_{n,t'}$            \\ \hline
$I^{Re}_{i,t'}$ & $-\frac{V_{i,t'}^{Re}}{V_{i,t'}^{2}}$  & $\frac{V_{i,t'}^{Re}}{V_{i,t'}^{2}}$ & $-\frac{V_{i,t'}^{Re}}{V_{i,t'}^{2}}$  & $\frac{V_{i,t'}^{Im}}{V_{i,t'}^{2}}$   & $\frac{V_{i,t'}^{Im}}{V_{i,t'}^{2}}$   & $-\frac{V_{i,t'}^{Im}}{V_{i,t'}^{2}}$ \\ \hline
$I^{Im}_{i,t'}$ & $- \frac{V_{i,t'}^{Im}}{V_{i,t'}^{2}}$ & $\frac{V_{i,t'}^{Im}}{V_{i,t'}^{2}}$ & $- \frac{V_{i,t'}^{Im}}{V_{i,t'}^{2}}$ & $- \frac{V_{i,t'}^{Re}}{V_{i,t'}^{2}}$ & $- \frac{V_{i,t'}^{Re}}{V_{i,t'}^{2}}$ & $\frac{V_{i,t'}^{Re}}{V_{i,t'}^{2}}$  \\ \hline\hline
\end{tabular}
\end{table}

\textbf{Step 2 - } Using $\partial \pmb{I^{Re}}/\partial \pmb{a_n}$ and $\partial \pmb{I^{Im}}/\partial \pmb{a_n}$ from Step 1 (Table \ref{table_Sen}), $\partial \pmb{V^{Re}}/\partial \pmb{a}$ and $\partial \pmb{V^{Im}}/\partial \pmb{a}$ are obtained employing the network-wide relationship between nodal voltages and current injections:
\begin{equation}   \label{eq:Sen_eq1} 
\left[
\begin{array}{ccc}
\frac{\partial \pmb{V^{Re}}}{\partial \pmb{a_n}}  \\
\frac{\partial \pmb{V^{Im}}}{\partial \pmb{a_n}}
\end{array}
\right] 
=
\left[
\begin{array}{ccc}
Y^{11} & Y^{12} \\
Y^{21} & Y^{22} 
\end{array}
\right]^{-1}
\left[
\begin{array}{ccc}
\frac{\partial \pmb{I^{Re}}}{\partial \pmb{a_n}}  \\
\frac{\partial \pmb{I^{Im}}}{\partial \pmb{a_n}}
\end{array}
\right] 
\end{equation}
where, the modified network bus admittance sub-matrices are determined as follows:
\begin{equation}
Y^{11} = Y^{Re}-Y^{(Re,Re)}_{D}, Y^{12} = -Y^{Im}-Y^{(Re,Im)}_{D}\label{eq_Sen1}
\end{equation}
\begin{equation}
Y^{21} = Y^{Im}-Y^{(Im,Re)}_{D},Y^{22} = Y^{Re}-Y^{(Im,Im)}_{D}\label{eq_Sen2}
\end{equation}
here, $Y^{Re}$ and $Y^{Im}$ are the real and imaginary parts of the original bus admittance matrix. The elements in diagonal matrices $Y^{(Re,Re)}_{D}$, $Y^{(Re,Im)}_{D}$, $Y^{(Im,Re)}_{D}$ and $Y^{(Im,Im)}_{D}$ are calculated using the following equations \cite{Sen_2016}:
\begin{equation}
Y^{(Re,Re)}_{D}(i,i) = \frac{p_{i,n,t'}}{V_{i,t'}^{2}}-\frac{2V_{i,t'}^{Re}(p_{i,n,t'}V_{i,t'}^{Re}+ q_{i,n,t'}V_{i,t'}^{Im})}{V_{i,t'}^{4}}\label{eq_Sen_YD1}
\end{equation}
\begin{equation}
Y^{(Re,Im)}_{D}(i,i) = \frac{q_{i,n,t'}}{V_{i,t'}^{2}}-\frac{2V_{i,t'}^{Im}(p_{i,n,t'}V_{i,t'}^{Re}+ q_{i,n,t'}V_{i,t'}^{Im})}{V_{i,t'}^{4}}\label{eq_Sen_YD2}
\end{equation}
\begin{equation}
Y^{(Im,Re)}_{D}(i,i) = -\frac{q_{i,n,t'}}{V_{i,t'}^{2}}-\frac{2V_{i,t'}^{Re}(p_{i,n,t'}V_{i,t'}^{Im}- q_{i,n,t'}V_{i,t'}^{Re})}{V_{i,t'}^{4}}\label{eq_Sen_YD3}
\end{equation}
\begin{equation}
Y^{(Im,Im)}_{D}(i,i) = \frac{p_{i,n,t'}}{V_{i,t'}^{2}}-\frac{2V_{i,t'}^{Im}(p_{i,n,t'}V_{i,t'}^{Im}- q_{i,n,t'}V_{i,t'}^{Re})}{V_{i,t'}^{4}}\label{eq_Sen_YD4}
\end{equation}

\textbf{Step 3 - } Noting that the current flow constraint returns and the rewards are also functions of branch current flows, the gradients of branch current flows are required to obtain $\partial J_{R_n}/\partial \pmb{a_n}$ and $\partial J_{C_m}/\partial \pmb{a_n}$. Using the branch current flow equations, these gradients are determined as a function of the derivatives of nodal voltages and current injections, as follows:
\begin{equation}
\frac{\partial I^{Re}_{ij,t'}}{\partial \pmb{a_{n,t'}}}=y_{ij}^{Im}(\frac{\partial V^{Im}_{i,t'}}{\partial \pmb{a_{n,t'}}}-\frac{\partial V^{Im}_{j,t'}}{\partial \pmb{a_{n,t'}}})-y_{ij}^{Re}(\frac{\partial V^{Re}_{i,t'}}{\partial \pmb{a_{n,t'}}}-\frac{\partial V^{Re}_{j,t'}}{\partial \pmb{a_{n,t'}}})\label{eq_Ire_IJ}
\end{equation}
\begin{equation}
\frac{\partial I^{Im}_{ij,t'}}{\partial \pmb{a_{n,t'}}}=y_{ij}^{Im}(\frac{\partial V^{Re}_{i,t'}}{\partial \pmb{a_{n,t'}}}-\frac{\partial V^{Re}_{j,t'}}{\partial \pmb{a_{n,t'}}})+y_{ij}^{Re}(\frac{\partial V^{Im}_{i,t'}}{\partial \pmb{a_{n,t'}}}-\frac{\partial V^{Im}_{j,t'}}{\partial \pmb{a_{n,t'}}})\label{eq_Iim_IJ}
\end{equation}
where, $I^{Re}_{ij}$ and $I^{Im}_{ij}$ are the real and imaginary parts of branch currents, $y^{Re}_{ij}$ and $y^{Im}_{ij}$ are the real and imaginary parts of branch admittance. 

\textbf{Step 4 - } Finally, using the derivatives obtained from Steps 1, 2, and 3, $\partial J_{R_n}/\partial \pmb{a_n}$ and $\partial J_{C_m}/\partial \pmb{a_n}$ are determined through straightforward algebraic manipulations. As an example, the gradient of reward function w.r.t. $P^{DG}_{n,t'}$ is calculated as:
\begin{equation}
\frac{\partial J_{R_n}}{\partial P^{DG}_{n,t'}}=\sum_{t'=t}^{t+T}(\lambda^{F}_{i,n}(2a_f+b_f)-\lambda^{R}_{n}\frac{\partial P^{PCC}_{n,t'}}{\partial P^{DG}_{n,t'}})\label{eq_Sen1_1}
\end{equation}
where, $\partial P^{PCC}_{n,t'}/\partial P^{DG}_{n,t'}$ is obtained using the outcomes of Steps 2 and 3, as follows:  
\begin{align}
\label{eq_SenPCC_1}
&\frac{\partial P^{PCC}_{n,t'}}{\partial P^{DG}_{n,t'}}= \frac{\partial V^{Re}_{i,t'}}{\partial P^{DG}_{n,t'}} I^{Re}_{ij,t'} + V^{Re}_{i,t'}\frac{\partial I^{Re}_{ij,t'}}{\partial P^{DG}_{n,t'}}\nonumber\\
&+\frac{\partial V^{Im}_{i,t'}}{\partial P^{DG}_{n,t'}} I^{Im}_{ij,t'}+V^{Im}_{i,t'}\frac{\partial I^{Im}_{ij,t'}}{\partial P^{DG}_{n,t'}}
\end{align}

Furthermore, $\partial J_{C_m}/\partial \pmb{a_n}$ for the global constraints \eqref{eqPF_v} and \eqref{eqPF_i} can be calculated using the outcomes of Steps 2 and 3:
\begin{equation}
\frac{\partial V_{i,t'}}{\partial \pmb{a_{n,t'}}}=\frac{V_{i,t'}^{Re}}{V_{i,t'}}\frac{\partial V^{Re}_{i,t'}}{\partial \pmb{a_{n,t'}}}+\frac{V_{i,t'}^{Im}}{V_{i,t'}}\frac{\partial V^{Im}_{i,t'}}{\partial \pmb{a_{n,t'}}}\label{eq_Sen11_1}
\end{equation}
\begin{equation}
\frac{\partial I_{ij,t'}}{\partial \pmb{a_{n,t'}}}=\frac{I_{ij,t'}^{Re}}{I_{ij,t'}}\frac{\partial I^{Re}_{ij,t'}}{\partial \pmb{a_{n,t'}}}+\frac{I_{ij,t'}^{Im}}{I_{ij,t'}}\frac{\partial I^{Im}_{ij,t'}}{\partial \pmb{a_{n,t'}}}\label{eq_Sen12_1}
\end{equation}

As can be seen in \eqref{eq3_3}-\eqref{eq3_essq}, the local constraint returns are trivial functions of the control actions. For example, the constraint return value for \eqref{eq3_3} is  $J_{C_{{5},t'}} = P^{DG}_{n,t'}$ which induces a simple gradient element w.r.t. control action $P^{DG}_{n,t'}$:  
\begin{equation}
\frac{\partial J_{C_{{5},t'}}}{\partial P^{DG}_{n,t'}}=1\label{eq_local_con}
\end{equation}

The gradients of constraint returns w.r.t. control actions for the remaining local constraints, \eqref{eq3_3q}-\eqref{eq3_essq}, can be obtained in a similar way.

\section{Derivation of $\partial \pmb{a_n}/\partial \pi_n$, $\partial \pi_n/\partial \pmb{\mu_n}$ and $\partial \pi_n/\partial \Sigma_n$}\label{app:proof}
 $\partial \pmb{a_n}/\partial \pi_n$, $\partial \pi_n/\partial \pmb{\mu_n}$ and $\partial \pi_n/\partial \Sigma_n$ are obtained using the probability density function of (D-dimensional) multivariate Gaussian distribution \cite{Math_2008}, which has the following general formulation:
\begin{equation}
f(\pmb{x};\pmb{\mu},\Sigma)=\frac{1}{\sqrt{|\Sigma|(2\pi)^D}}e^{-\frac{1}{2}(\pmb{x}-\pmb{\mu})^\top\Sigma^{-1}(\pmb{x}-\pmb{\mu})}\label{eq2_12_2}
\end{equation}
where $\pmb{x}$ is a random vector. To derive the gradients, first, the log-likelihood function of this multivariate Gaussian distribution \eqref{eq2_12_2} is obtained as follows:
\begin{equation}
L=\ln(f)=\ln{\frac{1}{\sqrt{|\Sigma|(2\pi)^D}}}-\frac{1}{2}(\pmb{x}-\pmb{\mu})^\top\Sigma^{-1}(\pmb{x}-\pmb{\mu})\label{eqA_1}
\end{equation}

The derivative of $L$ w.r.t. mean vector $\pmb{\mu}$ and covariance matrix $\Sigma$ can be written as follows:
\begin{multline}\label{eqA_3}
\frac{\partial L}{\partial \pmb{\mu}}=-\frac{1}{2}\frac{\partial (\pmb{x}-\pmb{\mu})^\top\Sigma^{-1}(\pmb{x}-\pmb{\mu})}{\partial \pmb{\mu}}\\
=-\frac{1}{2}(-2\Sigma^{-1}(\pmb{x}-\pmb{\mu}))=\Sigma^{-1}(\pmb{x}-\pmb{\mu})
\end{multline}

\begin{multline}\label{eqA_6}
\frac{\partial L}{\partial \Sigma}=-\frac{1}{2}(\frac{\partial \ln(|\Sigma|)}{\partial \Sigma}+\frac{\partial (\pmb{x}-\pmb{\mu})^\top\Sigma^{-1}(\pmb{x}-\pmb{\mu})}{\partial \Sigma})\\
=-\frac{1}{2}(\Sigma^{-1}-\Sigma^{-1}(\pmb{x}-\pmb{\mu})(\pmb{x}-\pmb{\mu})^\top\Sigma^{-1})
\end{multline}

Thus, using \eqref{eqA_3} and \eqref{eqA_6}, the derivatives of the function $f$ w.r.t. $\pmb{\mu}$ and $\Sigma$ can be shown in \eqref{eqA_7} and \eqref{eqA_8}, respectively: 
\begin{equation}
\frac{\partial f}{\partial \pmb{\mu}}=\frac{\Sigma^{-1}(\pmb{x}-\pmb{\mu})}{\sqrt{|\Sigma|(2\pi)^D}}e^{-\frac{1}{2}A}\label{eqA_7}
\end{equation}
\begin{equation}
\frac{\partial f}{\partial \Sigma}=-\frac{1}{2}\frac{(\Sigma^{-1}-\Sigma^{-1}(\pmb{x}-\pmb{\mu})(\pmb{x}-\pmb{\mu})^\top\Sigma^{-1})}{\sqrt{|\Sigma|(2\pi)^D}}e^{-\frac{1}{2}A}\label{eqA_8}
\end{equation}
where $A=(\pmb{x}-\pmb{\mu})^\top\Sigma^{-1}(\pmb{x}-\pmb{\mu})$. Similarly, the derivative of the function $f$ w.r.t. $\pmb{x}$ is shown as follows:
\begin{equation}
\frac{\partial f}{\partial \pmb{x}}=-\frac{\Sigma^{-1}(\pmb{x}-\pmb{\mu})}{\sqrt{|\Sigma|(2\pi)^D}}e^{-\frac{1}{2}A}\label{eqA_9}
\end{equation}

% or


% % you can choose not to have a title for an appendix
% % if you want by leaving the argument blank
% \section{}



\ifCLASSOPTIONcaptionsoff
  \newpage
\fi



\bibliographystyle{IEEEtran}
\bibliography{IEEEabrv,./bibtex/bib/IEEEexample}

\begin{IEEEbiography}[{\includegraphics[width=1in,height=1.25in,clip,keepaspectratio]{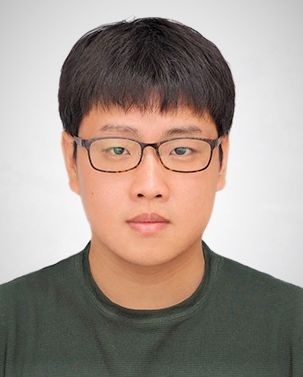}}]{Qianzhi Zhang}(S'17) is currently pursuing his Ph.D. in the Department of Electrical and Computer Engineering, Iowa State University, Ames, IA. He received his M.S. in electrical and computer engineering from Arizona State University in 2015. He has worked with Huadian Electric Power Research Institute from 2015 to 2016 as a research engineer. His research interests include the applications of machine learning and optimization techniques in power system operation and control.
\end{IEEEbiography}
	
\begin{IEEEbiography}[{\includegraphics[width=1in,height=1.25in,clip,keepaspectratio]{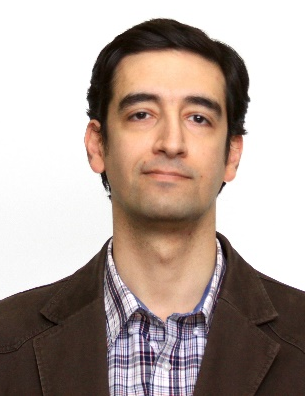}}]{Kaveh Dehghanpour}(S'14--M'17) received his B.Sc. and M.S. from University of Tehran in electrical and computer engineering, in 2011 and 2013, respectively. He received his Ph.D. in electrical engineering from Montana State University in 2017. He was a postdoctoral research associate at Iowa State University. His research interests include the applications of machine learning and data-driven techniques in power system monitoring and control.
\end{IEEEbiography}

\begin{IEEEbiography}[{\includegraphics[width=1in,height=1.25in,clip,keepaspectratio]{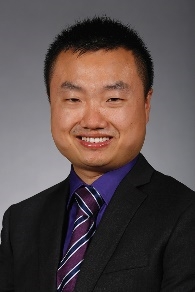}}]{Zhaoyu Wang}(S'13--M'15--SM’20) is the Harpole-Pentair Assistant Professor with Iowa State University. He received the B.S. and M.S. degrees in electrical engineering from Shanghai Jiaotong University in 2009 and 2012, respectively, and the M.S. and Ph.D. degrees in electrical and computer engineering from Georgia Institute of Technology in 2012 and 2015, respectively. His research interests include power distribution systems and microgrids, particularly on their data analytics and optimization. Dr. Wang is the Secretary of IEEE Power and Energy Society (PES) Award Subcommittee, Co-Vice Chair of PES Distribution System Operation and Planning Subcommittee, and Vice Chair of PES Task Force on Advances in Natural Disaster Mitigation Methods. He is an associate editor of IEEE Transactions on Power Systems, IEEE Transactions on Smart Grid, IEEE PES Letters, IEEE Open Access Journal of Power and Energy, and IET Smart Grid. Dr. Wang was the recipient of 2020 IEEE PES Outstanding Young Engineer Award.
\end{IEEEbiography}

\begin{IEEEbiography}[{\includegraphics[width=1in,height=1.25in,clip,keepaspectratio]{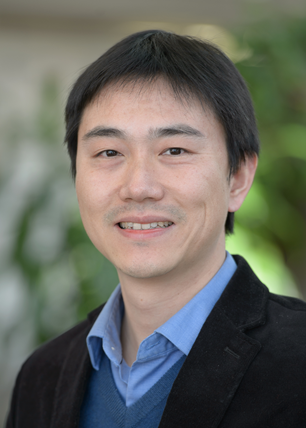}}]{Feng Qiu}(M'14--SM’18) received his Ph.D. from the School of Industrial and Systems Engineering at the Georgia Institute of Technology in 2013. He is a principal computational scientist with the Energy Systems Division at Argonne National Laboratory, Argonne, IL, USA. His current research interests include optimization in power system operations, electricity markets, and power grid resilience.
\end{IEEEbiography}

\begin{IEEEbiography}[{\includegraphics[width=1in,height=1.25in,clip,keepaspectratio]{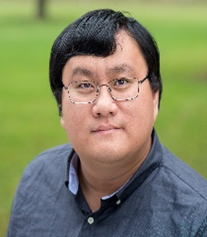}}]{Dongbo Zhao}(SM’16) received his B.S. degrees from Tsinghua University, Beijing, China, the M.S. degree from Texas A\&M University, College Station, Texas, and the Ph.D degree from Georgia Institute of Technology, Atlanta, Georgia, all in electrical engineering. He has worked with Eaton Corporation from 2014 to 2016 as a Lead Engineer in its Corporate Research and Technology Division, and with ABB in its US Corporate Research Center from 2010 to 2011. Currently he is a Principal Energy System Scientist with Argonne National Laboratory, Lemont, IL. He is also an Institute Fellow of Northwestern Argonne Institute of Science and Engineering of Northwestern University. His research interests include power system control, protection, reliability analysis, transmission and distribution automation, and electric market optimization. 

Dr. Zhao is a Senior Member of IEEE, and a member of IEEE PES, IAS and IES Societies. He is the editor of IEEE Transactions on Power Delivery, IEEE Transactions on Sustainable Energy, and IEEE Power Engineering Letters. He is the subject editor of subject “Power system operation and planning with renewable power generation” of IET Renewable Power Generation and the Associate Editor of IEEE Access.
\end{IEEEbiography}
\end{document}